\documentclass[a4paper,doc,natbib]{apa7}

\usepackage{amsmath,amssymb,amsfonts,amsthm}
\usepackage{graphicx}
\usepackage{natbib}
\usepackage{nicefrac}
\usepackage{float}
\usepackage{booktabs}
\usepackage{pdflscape}

\usepackage[ruled,vlined]{algorithm2e}

\SetAlgoNlRelativeSize{-1} % slightly smaller line numbers
\SetAlgoNlRelativeSize{0}
\SetNlSty{textbf}{}{}      % bold line numbers
\SetAlgoNlRelativeSize{-1}
\SetAlgoNlRelativeSize{0}
\SetKwInput{KwIn}{Input}
\SetKwInput{KwOut}{Output}

\newcommand{\prob}[1]{\mathbb{P}\left({#1}\right)}

\newcommand{\prior}[1]{\pi\left({#1}\right)}

\newcommand{\BF}[1]{\mathrm{BF}_{#1}}
\newcommand{\rhat}{\hat{R}}

\newcommand{\dnorm}[1]{\mathrm{Normal}\left({#1}\right)}
\newcommand{\dinvgamma}[1]{\mathrm{InverseGamma}\left({#1}\right)}

\newcommand{\predictor}{\beta}
\newcommand{\predictors}{\bm{\predictor}}
\newcommand{\indicator}{\gamma}
\newcommand{\indicators}{\bm{\indicator}}

\title{Reversible Jump MCMC With No Regrets: Bayesian Variable Selection Using Mixtures of Mutually Singular Distributions}

\authorsnames[1,2,3,1]{D. {van den Bergh}, M. A. Clyde, A. E. Raftery, M. Marsman}
\authorsaffiliations{{Department of Psychology, University of Amsterdam},
{Department of Statistical Science, Duke University},
{Department of Statistics, University of Washington}}
\authornote{Correspondence concerning this article should be addressed to:~\\Don van den Bergh~\\University of Amsterdam, Psychological Methods~\\Nieuwe Achtergracht 129B ~\\PO Box 15906, 1001 NK Amsterdam, The Netherlands~\\E-mail may be sent to d.vandenbergh@uva.nl.}

\shorttitle{Reversible Jump With No Regrets}

\date{}

\abstract{Bayesian variable selection requires sampling from a posterior distribution that combines discrete model indicators with continuously varying parameters, a challenge often addressed through reversible jump Markov chain Monte Carlo (RJMCMC). Despite its generality, RJMCMC is widely regarded as difficult to design and implement correctly. We present mixtures of mutually singular (MoMS) distributions as a transparent alternative in which competing models are represented within a single fixed-dimensional parameter space partitioned into mutually singular subspaces. We show that this formulation reproduces the exact spike-and-slab interpretation of Bayesian variable selection and that, under appropriate constructions, MoMS and RJMCMC share the same Metropolis--Hastings acceptance probability. On a benchmark dataset with ten predictors, both methods recover posterior inclusion probabilities that match full enumeration, while MoMS achieves comparable or superior effective sample size per second relative to a carefully engineered RJMCMC scheme. We further illustrate the approach in a mixed-effects logistic regression for a sleep-and-memory experiment and in factor-loading selection for a multidimensional generalized partial credit model. Together, these results show that Bayesian variable selection can be carried out within standard fixed-dimensional Markov chain Monte Carlo methodology---without regret.}


\begin{document}

\maketitle

\section{Introduction}

% Context of the specific journal
Bayesian statistics is becoming increasingly popular in psychometrics \citep{zagaria2024bayesian, PfadtEtAl2025}. This popularity stems from the suitability of Bayesian methods for complex psychometric models and their coherent treatment of inferential uncertainty.

Uncertainty in psychometric models concerns both the values of model parameters and the structure of the model \citep{Kaplan_2021, ClydeGeorge_2004}. For example, researchers may be uncertain about which predictor variables to include in a regression model, which item-factor loadings to include in a factor model, or which edges to include in a graphical model. In each of these cases, every specific selection or configuration of effects corresponds to a distinct statistical model.

Model uncertainty especially impacts inference of complex psychometric models, such as multidimensional latent variable models, and graphical models. For example, a Bayesian analysis by \citet{HuthEtAl_inpress_evidence} of psychometric networks from 126 published studies found substantial uncertainty about which edges to include or exclude. Two common Bayesian approaches to dealing with this type of uncertainty are model comparison, which evaluates the relative plausibility of competing models, and variable selection, which aims to identify effects that have high predictive value.

In Bayesian model comparison, model uncertainty is addressed by considering a finite set of competing models.
% \footnote{The so called ``$\mathcal{M}$-closed'' situation \citep{bernardoBayesianTheory1994}} 
% Don: I added the footnote above at first, but removed it later again. I think this is a can of worms that we shouldn't open. The M-open vs. M-closed discussion is somewhat of a tangent to the paper because then we quickly get into stacking and whatnot. Furthermore, we also don't strictly need to adhere to the M-closed view, BMA is defensible (and IMO sensible) in the M-open case as well.
Suppose we consider $M \in \mathbb{N}$ models and are interested in the relative plausibility of a given model $\mathcal{M}_m$ ($m = 1,\dots,M$) given the observed data $y$ and the prior model probabilities. This relative plausibility is quantified by the posterior model probability:
\[
p(\mathcal{M}_m \mid y) = \frac{p(y \mid \mathcal{M}_m)\, p(\mathcal{M}_m)}{\sum_{m^\prime =1}^M\, p(y \mid \mathcal{M}_{m^\prime} )\, p(\mathcal{M}_{m^\prime} )}.
\]

In practice, calculating posterior model probabilities is subject to two limitations. First, model comparison requires evaluation of the marginal likelihood for each model,
\[
p(y \mid \mathcal{M}_m) = \int p(y \mid \theta,\mathcal{M}_m) \, p(\theta \mid \mathcal{M}_m) \,\text{d}\theta,
\]
where $\theta$ denotes the model-specific parameters. Closed-form expressions for the marginal likelihood are available only in restrictive cases, and for more complex models it must be approximated using numerical methods such as Laplace approximations \citep{TierneyEtAl_1989}, importance sampling \citep{kass1995bayes}, bridge sampling \citep{gronau2017tutorial, meng1996simulating}, reciprocal importance sampling \citep{MetodievEtal2025}, or thermodynamic integration \citep{lartillot2006computing}; see \citet{llorente2023marginal} for an extensive review. Because many of these approaches are computationally intensive, they are feasible only when the number of competing models $M$ is relatively small.
Second, even when marginal likelihoods can be computed or approximated, the number of competing models can be prohibitively large. For example, a graphical model with $p = 10$ variables already admits more than 35 trillion distinct network structures, making exhaustive evaluation infeasible even under idealized computational assumptions.

In Bayesian variable selection, we consider $E \in \mathbb{N}$ possible effects and define an indicator vector $\indicators_m \in \{0,1\}^E$ of length $E$, where the $e$-th element indicates whether effect $e$ is included in model $m$ (i.e., $\indicator_{me} = 1$) or excluded (i.e., $\indicator_{me} = 0$). In this setting, models $\mathcal{M}_m$ are implicitly defined by different configurations of included and excluded effects $\indicators_m$ ($m = 1,\dots,M = 2^E$).
The posterior inclusion probability for an effect $e$ is obtained by aggregating posterior model probabilities,
\[
p(\indicator_e = 1 \mid y) = \sum_{m=1}^M \, \mathcal{I}(\indicator_{me} = 1)\, p(\mathcal{M}_m\mid y).
\]

Direct evaluation of posterior inclusion probabilities is hindered by the need to compute marginal likelihoods and to evaluate them for all possible models. Bayesian variable selection addresses these limitations by treating the vector of effect indicators $\indicators_m$ as a parameter of the model and jointly sampling the indicators and model parameters using Markov chain Monte Carlo (MCMC). Rather than enumerating all models and evaluating their marginal likelihoods, posterior model- and inclusion probabilities can be approximated using Monte Carlo averages over the sampled indicator configurations.

%RJMCMC
A major complication in using MCMC to simulate from the posterior distribution of effect indicators and model parameters is that the dimension of the model parameters may change as one moves between models. Standard MCMC algorithms assume that the dimension of the posterior distribution is fixed. Transdimensional MCMC relaxes the fixed dimensionality assumption and can simulate from the posterior distribution of effect indicators and model parameters. By far the most popular such approach is reversible jump MCMC (RJMCMC) proposed by \citet{Green_1995}.

Green generalized the Metropolis-Hastings algorithm to allow for between-model ``jumps'' while mapping the parameters of these models onto each other and ensuring that \textit{detailed balance} is achieved, i.e., the between-model jump is reversible. While RJMCMC allows for changes in parameter dimensions, it is so flexible that its design is considered very complicated. Folklore has it that no one has ever designed and implemented RJMCMC and not regretted it.

%SSVS
The practical difficulty of using RJMCMC for variable selection has motivated the development of simpler alternatives that avoid transdimensional moves. One particularly influential alternative is the continuous spike-and-slab prior, originally proposed by \citet{GeorgeMcCulloch_1993}. Rather than allowing parameters to enter or leave the model, the original continuous spike-and-slab approach represents each effect as arising from a mixture of two Gaussian distributions: a “slab” distribution with relatively large variance that permits substantial effects, and a “spike” distribution with very small variance that concentrates mass near zero.

Because effects are never exactly zero under a continuous spike-and-slab prior, the dimension of the parameter space remains fixed across all configurations, so standard MCMC algorithms can be used. This computational convenience has led to widespread adoption of the approach, including for predictor selection in linear regression \citep{GeorgeMcCulloch_1993}, item–factor selection \citep{MavridisNtzoufras_2014}, and edge selection in both continuous and discrete graphical models \citep{Wang_2015, ParkJinSchweinberger_2022, MarsmanEtAl_2022_rbinnet}.

However, this convenience comes at a cost. The performance of continuous spike-and-slab models depends critically on the specification of the spike and slab variances. When these variances are fixed, Gaussian spike-and-slab priors are not model-selection consistent: recovery of the true model is not guaranteed as the sample size grows \citep{LyWagenmakers_2022, NarisettyHe_2014, MarsmanEtAl_2022_rbinnet}. Consistency requires spike and slab variances that scale with the sample size \citep[e.g.,][]{ishwaranSpikeSlabVariable2005, NarisettyHe_2014, MarsmanEtAl_2022_rbinnet}, which fundamentally changes the behavior of the prior distribution and introduces additional tuning decisions. 
Although approaches such as the spike-and-slab LASSO mitigate these tuning challenges by adaptively learning the required scaling \citep{rockovaSpikeandSlabLASSO2018, baiSpikeandslabMeetsLASSO2021}, inference is still carried out over the full, fixed-dimension parameter space. As a result, computational and memory demands can become substantial in high-dimensional settings, and empirical comparisons indicate that continuous spike-and-slab approaches need not match the performance of leading Bayesian methods based on discrete model spaces \citep{porwal2022comparing}.
Despite its practical appeal, the continuous spike-and-slab prior therefore does not fully resolve the tension between theoretical guarantees and practical simplicity that motivates alternatives to RJMCMC.

%MoMs
The problem lies not in the spike-and-slab priors, but in how they encode model structure. Mixtures of mutually singular (MoMS) distributions define a discrete spike-and-slab approach that models the selection of effects by mixing over mutually singular parameter subspaces: one in which an effect is fixed exactly at zero, and a complementary subspace in which it is free to vary. In contrast to continuous spike-and-slab approaches, variable selection is expressed as moving between these distinct subspaces rather than as gradual shrinkage toward zero.

This paper brings MoMS distributions to the attention of psychometricians and Bayesian statisticians working on variable selection problems.
Originally introduced by \citet{GottardoRaftery_2008}, building on earlier geometric formulations of transdimensional MCMC \citep{PetrisTardella_2003}, the MoMS framework has received comparatively little attention relative to RJMCMC and continuous spike-and-slab approaches.
This is despite a growing body of methodological and applied work \citep[e.g.,][]{OhKim_2010_moms, SavitskyEtAl_2011_moms, LevineEtAl_2012_moms, Oh_2012_moms,  StingoEtAl_2013_moms, RosselEtAl_2013_moms, PetersonEtAl_2015_moms, BoggisMiloWalters_2016_moms, EdefontiParmigiani_2017_moms, WadsworthEtAl_2017_MoMSGibbs, ShaddoxEtAl_2018_moms, KolovskyVanucci_2020_MoMSAddDelete, CastellettiEtAl_2020_moms, GrimmetEtAl_2020_moms, MarsmanEtAl_2025_ordinal, AvalosEtAl_2025_moms, MarsmanEtAl_2025_ttest}.

Despite receiving little attention in applied Bayesian work, the MoMS formulation has important computational consequences. Because each subspace admits a fixed-dimensional parameterization, MoMS-based models use standard MCMC algorithms, without requiring transdimensional proposals. As a result, MoMS provides a straightforward alternative to RJMCMC for Bayesian variable selection. RJMCMC can be formulated so that it shares the same Metropolis–Hastings acceptance probability as MoMS-based algorithms, an equivalence previously noted by \citet[][Chapter 3]{Savitsky_2010_moms_equir_jmcmc} for Gaussian process priors. Despite this equivalence, the two approaches remain conceptually distinct in how model structure is represented. By framing variable selection directly in terms of mutually singular distributions, the resulting algorithms are easier to design, easier to reason about, and easier to implement correctly---without regret.

%Organization
The remainder of the paper is organized as follows.
In Section~\ref{sec:approaches}, we compare three approaches to Bayesian variable selection: enumeration, RJMCMC, and MoMS. We then study their numerical behavior on a diabetes dataset, where full enumeration is feasible and therefore provides a benchmark for posterior model probabilities and inclusion probabilities. Next, we apply the MoMS approach in two case studies: mixed-effects logistic regression and item factor selection in the multidimensional generalized partial credit model.
We conclude with a discussion of between-model moves, adaptive proposals, and practical recommendations for implementing custom MCMC algorithms.

\section{Approaches to Bayesian Variable Selection}\label{sec:approaches}
We introduce three approaches to Bayesian variable selection: enumeration of individual models, reversible jump MCMC (RJMCMC), and mixtures of mutually singular (MoMS) distributions. Each of these methods evaluates posterior model- and inclusion probabilities, but they differ substantially in their assumptions, computational demands, and ease of implementation. Detailed technical treatments and derivations can be found in  \citet{Green_1995}, and \citet{GottardoRaftery_2008}. We will provide a conceptual overview and illustrate the methods using a running example.

We use a linear regression model with a limited number of predictors, allowing marginal likelihoods to be computed directly and enumeration to serve as a benchmark. Enumeration is conceptually straightforward but relies on two strong assumptions: (i) the number of competing models $M$ is small enough to be tractable, and (ii) the marginal likelihoods $p(y \mid \mathcal{M}_m)$ are available in closed form or can be efficiently approximated. These assumptions rarely hold in high-dimensional settings, necessitating MCMC-based approaches. Such approaches construct a Markov chain whose stationary distribution is the joint posterior $p(\theta, \indicators \mid y)$. This yields a sequence of draws,
\[
(\theta^{(0)}, \indicators^{(0)}), (\theta^{(1)}, \indicators^{(1)}), (\theta^{(2)}, \indicators^{(2)}), \dots,
\]
and their Monte Carlo averages estimate posterior inclusion probabilities and model-averaged parameter estimates without enumerating all models or evaluating marginal likelihoods. For instance, the Monte Carlo estimate of the posterior inclusion probability for effect $e$ is
\[
\bar{\indicator}_e = \frac{1}{T}\sum_{t=1}^T \indicator_e^{(t)} \approx p(\indicator_e = 1 \mid y),
\]
and the corresponding model-averaged posterior mean is $\bar{\theta}_e$. (A certain number of the first draws are often discarded as warmup, to deal with the fact that MCMC samplers need some time to converge towards the stationary distribution.)
Code to reproduce the analyses is available at \url{https://github.com/vandenman/Supplement-Reversible-Jump-With-No-Regrets}.

\subsection{The Bayesian Linear Regression Model}\label{sec:running_example}

Suppose we observe an outcome variable $y = (y_1, \ldots, y_n)^T$ and a set of $p$ potential predictor variables collected in the design matrix $\bm{X} \in \mathbb{R}^{n \times p}$. The linear regression model assumes that each observation is generated as
\[
y_i = \mu + X_i^T \beta + \epsilon_i, \qquad i = 1, \ldots, n,
\]
where $\mu$ denotes the intercept, $\beta = (\beta_1, \ldots, \beta_p)^T$ are the regression coefficients, $X_i$ is the $i$-th row of $\bm{X}$ (assumed centered for convenience), and the error terms are independently normally distributed, $\epsilon_i \sim \mathcal{N}(0, \sigma^2)$. To complete the Bayesian specification, we place a prior on the model parameters.

We adopt the Jeffreys–Zellner–Siow (JZS) prior \citep{liang2008mixtures}. The JZS prior is a hierarchical extension of Zellner’s $g$-prior \citep{Zellner1980a}, and has the attractive property of yielding Bayes factors that are consistent and scale-invariant \citep{Jeffreys_1961, berger1998bayes}. Specifically, we use the improper Jeffreys’s prior $\pi(\mu, \sigma^2) \propto 1/\sigma^2$ for the intercept and error variance, and for the regression coefficients we assume
\begin{align*}
    \pi(\beta \mid \sigma^2, g, \mathcal{M}_j) &= \mathcal{N}\!\left(0,\, g\sigma^2 (\bm{X}_{\mathcal{M}_j}^T \bm{X}_{\mathcal{M}_j})^{-1}\right),\\
    \pi(g) &= \text{Inv-Gamma}\!\left(\tfrac{1}{2}, \tfrac{n}{2}\right),
\end{align*}
where $\bm{X}_{\mathcal{M}_j}$ denotes the design matrix restricted to the predictors included in model $\mathcal{M}_j$. Integrating out $g$ yields a multivariate Cauchy prior on $\beta$, which provides automatic shrinkage and guards against overfitting.

For the linear regression example, enumeration, RJMCMC, and MoMS are compared under this same model-specific JZS prior. Under model $\mathcal{M}_j$, the coefficients corresponding to the predictors included in $\mathcal{M}_j$ are assigned the multivariate JZS prior given above. The differences between the three methods are therefore computational rather than statistical: they explore the same posterior distribution under the same prior specification. 

\subsubsection{Analytic Bayes Factor}
Under the JZS prior, the Bayes factor comparing a specific model $\mathcal{M}_j$ to the null model can be expressed as a one-dimensional integral over $g$:
\begin{align}\label{eq:jzs_bf}
    \BF{j0} = \int_0^\infty
    \left(1 + g\right)^{(N - 1 - p_{j}) / 2}
    \left(1 + \left(1 - R_j^2\right)g\right)^{-(N-1) / 2}
    \,\,\prior{g} \enspace \mathrm{d}g ,
\end{align}
where $p_j$ denotes the number of predictors in model $\mathcal{M}_j$ and $R_j^2$ is the coefficient of determination. Because $\BF{j0}$ reduces to a one-dimensional integral over $g$, it remains straightforward to evaluate numerically even when $p_j$ is large. 
Implementations of the JZS prior and corresponding Bayes factors are available in the \texttt{BayesFactor} package \citep{morey2024Bayesfactor} and the \texttt{BAS} package \citep{clyde2025BAS} in R. These implementations are also used in JASP \citep{love2019jasp, bergh2021tutorial}.

% {\sc MAC NOTE:  are we using the JZS prior in all of the linear regression examples:  enumeration, MoMs, RJMCMC?}
% Don: Yes.
% Maarten: Adressed with the paragraph above.

\subsubsection{The Diabetes Data}
We use the diabetes dataset originally analyzed by \citet{efron2004least}, which has been widely used in Bayesian regression studies. The dataset contains measurements from $N = 442$ diabetes patients, with the outcome variable being a quantitative measure of disease progression one year after baseline. Ten predictors are available: age, sex, body mass index (BMI), average blood pressure (BP), and six blood serum measurements (S1–S6). The moderate number of predictors makes it feasible to enumerate models and compute posterior inclusion probabilities exactly, while still providing a realistic testbed for evaluating MCMC-based methods such as RJMCMC and MoMS.

\subsection{Enumeration}\label{sec:enumeration}
Enumeration is the most direct and transparent method of Bayesian variable selection. We evaluate the marginal likelihood of each candidate model and combine these with the prior model probabilities to compute the posterior inclusion probabilities exactly. This exhaustive approach establishes a gold-standard baseline against which Monte Carlo methods such as RJMCMC and MoMS can be evaluated.

\subsubsection{Computation}
For each model $\mathcal{M}_j$, the marginal likelihood $p(y \mid \mathcal{M}_j)$ is
computed using Eq.~\ref{eq:jzs_bf}. Assuming that $\mathcal{M}_1$ is the null model containing only the intercept, the posterior model probabilities follow as
\[
p(\mathcal{M}_j \mid y) = \frac{p(y \mid \mathcal{M}_j)\, p(\mathcal{M}_j)}{\sum_{m = 1}^M\, p(y \mid \mathcal{M}_{m} )\, p(\mathcal{M}_{m} )}=
\frac{\text{BF}_{j0} \, \frac{p(\mathcal{M}_j)}{p(\mathcal{M}_1)}}
     {1 + \sum_{m=2}^M \text{BF}_{m0} \, \frac{p(\mathcal{M}_m)}{p(\mathcal{M}_1)}},
\]
and posterior inclusion probabilities for predictor $e$ are obtained by summing over all models that include it:
\[
p(\indicator_e = 1 \mid y) =
\sum_{j: \indicator_e = 1} p(\mathcal{M}_j \mid y).
\]
Model‐averaged estimates such as
$\beta_{BMA} = \sum_j p(\mathcal{M}_j \mid y)
\,\mathbb{E}[\beta_e \mid y,\mathcal{M}_j]$
can be derived directly. We assign a uniform prior over the model space, $p(\mathcal{M}_j) = 1/M$ for $j = 1, \dots, M$.

\subsection{Reversible Jump Markov Chain Monte Carlo}\label{Sec::reversible_jump}

Reversible jump Markov chain Monte Carlo (RJMCMC) \citep{Green_1995} extends the Metropolis–Hastings algorithm to variable selection problems in which the dimension of the parameter vector depends on the model. Models are different configurations of the effect indicators $\indicators$. RJMCMC targets the joint posterior distribution over the effect indicators and within-model parameters,
\[
p(\indicators, \bm{\beta} \mid \bm{y}, \bm{X})
\propto
p(\bm{y} \mid \bm{X}, \bm{\beta}, \indicators)\,
\pi(\bm{\beta} \mid \indicators)\,
\pi(\indicators).
\]

Correctness of a Markov chain Monte Carlo algorithm requires that the target distribution be invariant under the transition mechanism. A sufficient condition for invariance is the \emph{detailed balance condition}, which requires that, under the target distribution, the probability of moving from one state to another is balanced by the probability of moving back (Appendix~\ref{App::detailed_balance}). Standard Metropolis--Hastings algorithms satisfy detailed balance by construction when proposals are defined on a fixed-dimensional parameter space (Appendix~\ref{App::metropolis_hastings}).

A practical difficulty in designing an RJMCMC sampler is that the forward and reverse proposals in variable selection problems are defined on parameter spaces of different dimension. When a predictor is added or removed, the corresponding parameter vectors live on spaces that cannot be compared directly. In Appendix~\ref{app:rjmcmc_details} we show that the detailed balance equation cannot be formulated for standard Metropolis--Hastings proposals, and the usual acceptance ratio therefore breaks down.

RJMCMC resolves this problem by modifying how between-model proposals are constructed so that the detailed balance condition is met. Rather than attempting to compare proposals defined on parameter spaces of different dimension, RJMCMC reformulates each between-model move on an augmented space of fixed dimension. As shown in Appendix~\ref{app:rjmcmc_details}, this reformulation restores the ability to write down and solve the detailed balance equation using standard Metropolis--Hastings arguments.

To construct augmented proposals, RJMCMC introduces auxiliary variables and defines a deterministic, one-to-one mapping between the parameters of the current and proposed models. Specifically, a between-model proposal is defined by a one-to-one transformation
\[
(\predictors^\star, \mathbf{u}^\star) = g(\predictors, \mathbf{u}),
\]
where the auxiliary variables $\mathbf{u}$ and $\mathbf{u}^\star$ ensure that the forward and reverse densities have the same support. The role of this transformation, the introduction of auxiliary variables, and the resulting form of the proposal densities are derived in Appendix~\ref{app:rjmcmc_details}.

For proposals formulated on the augmented space, the acceptance probability follows from the Metropolis solution of the detailed balance equation (Appendix~\ref{App::metropolis_hastings}). For a proposal based on the transformation $g$, the acceptance probability takes the general form
\begin{equation}\label{eq:acc_rjmcmc}
\alpha_{\text{RJ}} =
\min\!\left\{1,\frac{
p(\indicators^\star,\predictors^\star \mid \bm{y}, \bm{X})\,
q(\indicators^\star \mid \indicators) \,q(\mathbf{u}^\star \mid \predictors^\star,\indicators^\star,\indicators)
}{
p(\indicators,\predictors \mid \bm{y}, \bm{X})\,
q(\indicators \mid \indicators^\star) \,q(\mathbf{u} \mid \predictors,\indicators,\indicators^\star)
}\,
\left|\tfrac{\partial(\predictors^\star, \mathbf{u}^\star)}
{\partial(\predictors, \mathbf{u})}\right|\right\},
\end{equation}
Here, $(\predictors, \mathbf{u})$ denotes the current augmented state of the chain, and $(\predictors^\star, \mathbf{u}^\star)$ the proposed state. The determinant term is the Jacobian of the transformation $g$, accounting for the change of variables induced by the mapping between augmented parameterizations.

The distribution $q(\indicators^\star \mid \indicators)$ in Eq.~\ref{eq:acc_rjmcmc} specifies how proposed models are generated from the current model. In Bayesian variable selection, RJMCMC algorithms commonly apply local add or delete moves that modify the inclusion status of a single predictor at a time. When a predictor is added, a new regression coefficient must be proposed; when a predictor is removed, its coefficient must be eliminated. The efficiency of the sampler depends on how new coefficients are proposed and how existing coefficients are adjusted. Poorly chosen proposals lead to low acceptance rates and slow exploration of the model space.

Several principled constructions for add and delete proposals have been developed in the literature. \citet{forster2012reversible} introduced an add/delete scheme for generalized linear (mixed) models that yields efficient between-model moves by aligning proposed coefficients with the likelihood under the candidate model. In the following subsection, we describe this procedure for the linear regression model used throughout the paper.
% {\sc MAC
% NOTE: the add-delete (swap) MCMC algorithm in BAS is just a special case of RJMCMC but uses a different proposal so that there is cancellation of the terms involving $\beta$'s that results in the acceptance ratio being just the ratio of marginal likelihoods (with uniform priors on models) - see comment later for algorithm in \citet{Clyde_1999}.  The paper by \citet{DellaportasEtAl_2002}[sec 2.5] explicitly shows how a general fixed dimension MCMC algorithm that moves around the model space alone can be derived as a special case of RJMCMC when the proposal for $\beta_m$ is the posterior under model m (i.e linear models with priors like JZS or other mixtures of Gaussians, nice graphical models) worth mentioning as it connects the special case from BAS with the more general RJMCMC; similar comparison and other special cases in \citep{godsill_relationship_2001}. Both instances recast the pseudo-prior approach in \citet{CarlinChib_1995} as a special case of RJMCMC, which then simplifies to this in models where marginal likelihoods are available.  Could follow the next section - of course it does not generalize to more complex models but suggests that "independent" proposals for $\beta_m$ that are somehow "close" to the posterior would be efficient.  }

% Don: I addressed this further down where we compare RJMCMC and MOMS because there is a MOMS interpretation to these moves.

\subsubsection{Computation}\label{sec::rjcomputation}
This section describes the reversible jump sampler proposed by \citet{forster2012reversible}, specialized to the linear regression model, using add/delete proposals.

At iteration $t$ of the Markov chain, each between-model move proceeds by first selecting a predictor index $i \in \{1,\dots,p\}$. Conditional on the state of the selected indicator $\indicator_i^{(t)}$, we deterministically \emph{flip} this component,
\[
\indicator_i^\star = 1 - \indicator_i^{(t)},
\]
so that the proposed model differs from the current one only in the inclusion status of predictor $i$. This defines the candidate model $\indicators^\star$. When $\indicator_i^{(t)} = 0$ and $\indicator_i^\star = 1$, we attempt an \emph{add} move; when $\indicator_i^{(t)} = 1$ and $\indicator_i^\star = 0$, we attempt a \emph{delete} move.

Both add and delete moves are constructed using deterministic linear transformations between the current and proposed coefficient vectors, augmented by an auxiliary variable. In both cases, the transformation is triangular with unit determinant, so the Jacobian correction equals one and does not appear explicitly in the acceptance probabilities. The difference between add and delete moves is whether the auxiliary variable is generated stochastically or recovered deterministically.

We begin with the add move. The design matrix under the current model is $\bm{X}_{\indicators}$, and $\bm{s}_{\indicators^\star}$ denotes the column of the design matrix under the full model $\bm{X}_\ast$ that becomes active under the proposed model $\indicators^\star$. The proposal introduces a single auxiliary variable $u \in \mathbb{R}$ and forms the proposed coefficient vector through the linear transformation
\begin{equation}\label{eq:linear_transform_add}
\predictors_{\indicators^\star}
= g(\predictors_{\indicators},u)
=
\begin{pmatrix}
\textbf{I} &
-\,(\bm{X}_{\indicators}^{\top}\bm{X}_{\indicators})^{-1}
  \bm{X}_{\indicators}^{\top}\bm{s}_{\indicators^\star} \\
\mathbf{0} & 1
\end{pmatrix}
\begin{pmatrix}
\predictors_{\indicators} \\ u
\end{pmatrix}.
\end{equation}
Under this transformation, the coefficient of the added predictor is $u$, while the existing coefficients are adjusted according to the regression of the new column $\bm{s}_{\indicators^\star}$ on $\bm{X}_{\indicators}$. The transformation matrix is triangular with ones on the diagonal, so its determinant equals $1$ and the Jacobian does not contribute to the acceptance probability.

The auxiliary variable $u$ is proposed from a Gaussian density $\mathcal{N}(\mu, v)$ that uses information from the full model to inform the proposed coefficient. Following \citet{forster2012reversible}, the proposal parameters are defined using the residual vectors
\[
\mathbf{r}^{(s)}
= \bm{s}_{\indicators^\star}
- \bm{X}_{\indicators}
(\bm{X}_{\indicators}^{\top}\bm{X}_{\indicators})^{-1}
\bm{X}_{\indicators}^{\top}\bm{s}_{\indicators^\star},
\]
and
\[
\mathbf{r}^{(\eta)}
= \hat{\bm{\eta}}_{\mathcal{M}_\ast}
- \bm{X}_{\indicators}
(\bm{X}_{\indicators}^{\top}\bm{X}_{\indicators})^{-1}
\bm{X}_{\indicators}^{\top}\hat{\bm{\eta}}_{\mathcal{M}_\ast}.
\]
Here, $\hat{\bm{\eta}}_{\mathcal{M}_\ast} = \bm{X}_\ast \hat{\predictors}_\ast$ denotes the fitted values under the full model $\mathcal{M}_\ast$, evaluated at the posterior mode $\hat{\predictors}_\ast$.\footnote{Rather than fitting the full model, one could also use a different anchor such as the maximum likelihood estimate.} Using these quantities, the proposal variance and mean are
\begin{equation}\label{eq::forster_proposal_parameters}
v =
\left(
\hat{\sigma}_\ast^{-2}
\bm{s}_{\indicators^\star}^{\top}\mathbf{r}^{(s)}
\right)^{-1},
\qquad
\mu =
v\,
\hat{\sigma}_\ast^{-2}
\bm{s}_{\indicators^\star}^{\top}\mathbf{r}^{(\eta)},
\end{equation}
where $\hat{\sigma}_\ast^{2}$ is the residual variance under the full model $\mathcal{M}_\ast$. Thus, $v$ depends on the part of the new predictor that is not explained by the current model, while $\mu$ centers the proposal at values suggested by the full-model fit.

The acceptance probability for an add move follows from the general RJMCMC rule in Eq.~\eqref{eq:acc_rjmcmc} and is given in Algorithm~\ref{alg:rjmcmc_forster}.

For the delete move, no auxiliary variable is generated. Instead, the auxiliary value determined by the current coefficients is recovered through the inverse transformation
\begin{equation}\label{eq:linear_transform_del}
\begin{pmatrix}
\predictors_{\indicators^\star}\\
\mathbf{u}^\star
\end{pmatrix} = g^{-1}(\predictors_{\indicators}) = \begin{pmatrix}
\textbf{I} &
\,-(\bm{X}_{\indicators^\star}^{\top}\bm{X}_{\indicators^\star})^{-1}
  \bm{X}_{\indicators^\star}^{\top}\bm{s}_{\indicators^{\ast\ast}} \\
\mathbf{0} & 1
\end{pmatrix}
\, \predictors_{\indicators} ,
\end{equation}
where $\bm{s}_{\indicators^{\ast\ast}}$ denotes the column of the design matrix under the full model $\bm{X}_\ast$ that is deactivated under the proposed model, and $g^{-1}$ denotes the inverse of the transformation in Eq.~\eqref{eq:linear_transform_add}.

The delete-move acceptance probability again follows from Eq.~\eqref{eq:acc_rjmcmc} and is given explicitly in Algorithm~\ref{alg:rjmcmc_forster}.

Algorithm~\ref{alg:rjmcmc_forster} gives the full reversible jump sampler for the linear regression model using the add/delete proposals described above. For each iteration $t$ and each predictor $i$, the algorithm flips the inclusion indicator $\indicator_i$ to define a candidate model, proposes new regression coefficients using the linear transformation $g$ and the Gaussian proposal $q_{i}(u)$, and accepts or rejects the move according to the acceptance probability in Eq.~\ref{eq:acc_rjmcmc}. Within-model updates of $(\sigma^2,\predictors_{\indicators})$ are carried out in separate Gibbs or Metropolis–Hastings steps and are not specific to the reversible jump mechanism.

\begin{algorithm}
\caption{RJMCMC with Forster et al.\ (2012) add/delete proposals for the linear regression model}
\KwIn{Data $(\bm{X}, \bm{y})$, prior $\pi(\predictors,\indicators,\sigma^2,g)$ (see Section~\ref{sec:running_example}), total iterations $T$}
\KwIn{Full-model quantities: $\bm{\eta}_\ast = \bm{X}_\ast \hat{\predictors}_\ast$, $\hat{\sigma}_\ast^2$}
\KwOut{Samples $\{(\predictors^{(t)}, \indicators^{(t)})\}_{t=0}^T$ from $p(\predictors,\indicators \mid \bm{y}, \bm{X})$}

Initialize $(\predictors^{(0)}, \indicators^{(0)}, \sigma^{2(0)})$\;
\For{$t = 0$ \KwTo $T-1$}{
    Set $(\predictors, \indicators) \leftarrow (\predictors^{(t)}, \indicators^{(t)})$\;

    \For{$i = 1$ \KwTo $p$}{
        \tcp{Between–model add/delete proposal for predictor $i$}
        Form candidate inclusion vector $\indicators^\star$ by flipping $\indicator_i$\;
        Let $\bm{X}_{\indicators}$ be the current design matrix and
        let $\bm{s}_{\indicators^\star}$ be the column of $\bm{X}_\ast$ corresponding to predictor $i$\;
%        Compute the residual vectors $\mathbf{r}^{(s)}$ and $\mathbf{r}^{(\eta)}$\;
        Compute the proposal variance $v$ and mean $\mu$ in Eq.~\ref{eq::forster_proposal_parameters}\;
        \If{$\indicator_i^\star = 1$}{%
            \tcp{Add move}
            Draw $u \sim q_i(u \mid \indicators, \indicators^\star) = \mathcal{N}(\mu, v)$\;
            Use transformation $g$ in Eq.~\eqref{eq:linear_transform_add} to form the proposed coefficients $\predictors_{\indicators^\star}$\;
            %Insert zeros at excluded predictors to obtain the full vector $\predictors^\star$\;
            Compute the acceptance probability
            \[
            \alpha = \min\!\left\{1,\frac{p(\indicators^\star,\predictors^\star_{\indicators^\star} \mid \bm{y}, \bm{X}_{\indicators^\star})\,
                }{
                    p(\indicators,\predictors_{\indicators} \mid \bm{y}, \bm{X}_{\indicators})
                }\, \frac{1}{q_i(u \mid \indicators, \indicators^\star)}\right\};
            \]
        }
        \Else{%
            \tcp{Delete move}
            Use the inverse of the transformation in Eq.~\eqref{eq:linear_transform_add} to obtain the proposed coefficients under the smaller model, i.e.\ solve
            $(\predictors_{\indicators^\star}, u^\star) = g^{-1}(\predictors_{\indicators})$\;
            %Insert zeros at excluded predictors and predictor $i$ to obtain the full vector $\predictors^\star$\;
            Compute the acceptance probability
            \[
            \alpha = \min\!\left\{1,\frac{p(\indicators^\star,\predictors^\star_{\indicators^\star} \mid \bm{y}, \bm{X}_{\indicators^\star})\,}{p(\indicators,\predictors_{\indicators} \mid \bm{y}, \bm{X}_{\indicators})
                }\, \frac{q_{i}(u^\star \mid \indicators^\star, \indicators)}{1}\right\};
            \]
        }

        \tcp{Accept/reject between–model move}
        With probability $\alpha$, set $(\predictors_{\indicators}, \indicators) \leftarrow (\predictors^\star_{\indicators^\star}, \indicators^\star)$\;
    }

    Set $(\predictors^{(t+1)}, \indicators^{(t+1)}) \leftarrow (\predictors, \indicators)$\;

    \tcp{Within–model update (fixed $\indicators$)}
    Update $(\sigma^2, \predictors_{\indicators})$ in separate Gibbs steps\;
}
\label{alg:rjmcmc_forster}
\end{algorithm}

\subsection{Mixtures of Mutually Singular Distributions}\label{Sec:MoMS}

Reversible jump MCMC (RJMCMC) and the method of mixtures of mutually singular (MoMS) distributions both use Metropolis-Hastings to sample posterior distributions over models. A central difficulty in this setting is that different models are naturally associated with parameter spaces of different dimension, whereas the Metropolis-Hastings algorithm is defined on a fixed-dimensional space. The original RJMCMC approach of \citet{Green_1995} resolves this difficulty by augmenting states with auxiliary variables so that transitions between models can be evaluated within a common dimension. MoMS instead represents all models within a single fixed-dimensional space.

In Bayesian variable selection problems, some parameters can be exactly zero (e.g., excluded predictors), while others are continuous (e.g., included predictors). The familiar ``spike-and-slab'' priors, for example, expresses this idea as a mixture of a point mass at zero (the spike) and a continuous density (the slab),
\[
p(\predictor_i \mid \indicator_i)
= (1-\indicator_i)\,\mathbf{1}_{\{0\}}(\predictor_i)
+ \indicator_i\,p(\predictor_i)\,\mathbf{1}_{\mathbb{R}\setminus\{0\}}(\predictor_i),
\]
where the inclusion indicator $\indicator_i$ indicates whether effect $i$ is included. %Here, $\mathbf{1}_{A}(\predictor_i)$ denotes the indicator function that equals $1$ if $\predictior_i \in A$ and $0$ otherwise.
The two components of this mixture have disjoint support: the spike assigns all probability to the point $\predictor_i \in \{0\}$, whereas the slab assigns probability only to values $\predictor_i \in \mathbb{R}\setminus\{0\}$. Because their supports do not overlap, the components are \emph{mutually singular}.

This idea extends from a single coefficient to the full parameter vector. Each configuration of inclusion indicators then defines a distinct statistical model $\mathcal{M}_{\indicators}$, in which some coefficients are fixed to zero and others are free to vary continuously. Together, these models divide the joint parameter space into disjoint regions. Transitions between models move from one region to another within a fixed-dimensional space, rather than changing the space's dimension. %This formulation naturally generalizes to multivariate settings, where the slab component represents a joint prior distribution over the subset of included predictors.

% {\sc MAC NOTE: Is MoM using a product of independent priors with a discrete spike and slab - or is it using the same prior in JZS? - for example if all of the components of $\beta_{(-\gamma)} = \mathbf{0}$ have a multivariate spike with a continuous multivariate Cauchy "slab" then this is JVS;  The way the algorithms are portrayed it is not clear if we are using the same priors through-out for the linear regression example.  If we are using JZS for the non-zero component $\beta_\gamma$, then we should be explicit and provide notation that writes the prior as a mixture of a multivariate point mass at zero and the continuous JZS component. }
% Don: we use a multivariate prior, the usual JZS. I've mentioned in a sentence that this can be extended to multivariate slabs.
% Maarten: Adressed instead with the paragraph above.

MoMS redefines the state of the Markov chain so that it includes both the model indicators and the model parameters. The joint space consists of all coefficients $\predictors = (\predictor_1,\dots,\predictor_p)$ and their inclusion indicators $\indicators = (\indicator_1,\dots,\indicator_p)$:
\[
\mathcal{X} = \mathbb{R}^p \times \{0,1\}^p.
\]
Each model configuration $\indicators$ then corresponds to a distinct region of this joint space,
\[
\mathcal{S}_{\indicators}
= \bigl\{(\predictors,\indicators) :
\predictor_i = 0 \text{ if } \indicator_i = 0,\;
\predictor_i \in \mathbb{R}\setminus\{0\} \text{ if } \indicator_i = 1
\bigr\}.
\]
Different configurations of $\indicators$ define disjoint regions: $\mathcal{S}_{\indicators} \cap \mathcal{S}_{\indicators^\star} = \varnothing$. Two models can assign the same nonzero value to a common coefficient, but they still represent different points in the joint space because their indicator vectors differ. This construction keeps the overall state dimension fixed ---$p$ continuous coordinates plus $p$ binary coordinates--- while assigning each model its own non‐overlapping ``slice'' of the joint space~$\mathcal{X}$. The posterior is supported on a union of disjoint regions $\{\mathcal{S}_{\indicators}\}$ within a single joint space, so Metropolis–Hastings updates apply directly \citep{GottardoRaftery_2008}.

Detailed balance for MoMS follows the same logic as in the Metropolis-Hastings algorithm. Because forward and reverse moves have densities that are defined on the same fixed space, the acceptance probability is given directly by the standard Metropolis--Hastings ratio:
\begin{equation}\label{acc:MoMS}
\alpha_{\text{MoMS}} = \min\!\left\lbrace 1,\; \frac{p(\indicators^\star,\predictors^\star\mid\bm{y},\bm{X})\, q(\indicators,\predictors\mid\indicators^\star,\predictors^\star)}{p(\indicators,\predictors\mid\bm{y},\bm{X})\, q(\indicators^\star,\predictors^\star\mid\indicators,\predictors)}\right\rbrace.
\end{equation}
The proposal $q(\indicators^\star, \predictors^\star \mid \indicators, \predictors)$ must generate only valid points within the non-overlapping regions $\mathcal{S}_{\indicators}$ and $\mathcal{S}_{\indicators^\star}$.

Implementing MoMS requires specifying proposal mechanisms for moves between models and for the associated regression coefficients. Although MoMS avoids the explicit use of auxiliary variables, the efficiency of the sampler still depends on how regression coefficients are proposed when predictors enter or leave the model. Following \citet{andrieu2008tutorial} and \citet{roberts2009examples}, we adopt random-walk Metropolis proposals for the continuous coefficients, with proposal variances adapted during warmup as described in Appendix~\ref{App::adaptive_metropolis}; this choice applies broadly across statistical models.

\subsubsection{Computation}

We implement the MoMS sampler as a Metropolis--within--Gibbs algorithm on the joint state space \((\predictors,\indicators) \in \mathbb{R}^p \times \{0,1\}^p\). Between-model moves update the indicator vector by flipping a single component, as in the reversible jump sampler described in Section~\ref{sec::rjcomputation}.

Conditional on the proposed indicators \(\indicators^\star\), a coefficient vector \(\predictors^\star\) is generated as follows. If \(\indicator_i^\star = 0\), the coefficient is set to zero. If \(\indicator_i^\star = 1\), it is sampled from a continuous normal distribution centered at its current value,
\[
\predictor_i^\star \sim \mathcal{N}(\predictor_i,\, \tau_i^2),
\]
with support restricted to \(\mathbb{R}\setminus\{0\}\). All remaining coefficients are left unchanged. This construction mirrors the support of the spike-and-slab prior and ensures that proposed states lie in region \(\mathcal{S}_{\indicators^\star}\).

The corresponding proposal density for the updated coefficient is
\[
q(\predictor_i^\star \mid \indicator_i^\star, \predictor_i) = (1 - \indicator^\star)\mathbf{1}_{\{0\}}(\predictor_i^\star) + \indicator_i^\star\, q(\predictor_i^\star \mid \predictor_i)\, \mathbf{1}_{\mathbb{R}\setminus \{0\}}(\predictor_i^\star).
\]
In particular, when $\indicator^\star_i =0$ the proposal density equals one, and when $\indicator^\star_i = 1$ it reduces to the normal density evaluated at $\predictor_i^\star$.

The proposal variance \(\tau_i^2\) is adapted during warmup of the Markov chain
using a standard adaptive Metropolis scheme. Appendix~\ref{App::adaptive_metropolis} details the Robbins–Monro approach we adopt here.

\begin{algorithm}
\caption{MoMS algorithm with random walk proposals for the linear regression model }\label{alg:MoMS}
\KwIn{Data $(\bm{X}, \bm{y})$, prior $\pi(\predictors,\indicators)$, total iterations $T$}
\KwIn{Metropolis proposal standard deviations $\boldsymbol{\tau}$ (See Appendix~\ref{App::adaptive_metropolis}).}
\KwOut{Samples $\{(\predictors^{(t)}, \indicators^{(t)})\}_{t=1}^T$ from $p(\predictors, \indicators \mid \bm{y}, \bm{X})$}

Initialize $(\predictors^{(0)}, \indicators^{(0)})$\;
\For{$t = 0$ \KwTo $T-1$}{
    Set $(\predictors, \indicators) \leftarrow (\predictors^{(t)}, \indicators^{(t)})$\;

    \For{$i = 0$ \KwTo $p$} {
        \tcp{Between–model add/delete proposal for predictor $i$}
        Form candidate inclusion vector $\indicators^\star$ by flipping $\indicator_i$\;
        \If{$\indicator_i^\star = 1$}{
            \tcp{Add move}
            Draw $\predictor^\star \sim q\left(\predictor^\star \mid \predictor_i\right) = \mathcal{N}(\predictor^\star  \mid \predictor_i, \tau_i^2)$\;
            Compute the acceptance probability;
            \[
            \alpha = \min\!\left\lbrace 1,\;
            \frac{p(\indicators^\star,\predictors^\star\mid \bm{y}, \bm{X})}{ p(\indicators,\predictors \mid \bm{y}, \bm{X})}\, \frac{1}{ q\left(\predictor^\star_i \mid \predictor_i\right) }
            \right\rbrace ;
            \]
        }
        \If{$\indicator_i^\star = 0$}{
            \tcp{Delete move}
            Set $\predictor^\star = 0$\;
            Compute the acceptance probability;
            \[
            \alpha = \min\!\left\lbrace 1,\;
            \frac{p(\indicators^\star,\predictors^\star\mid \bm{y}, \bm{X})}{ p(\indicators,\predictors \mid \bm{y}, \bm{X})}\, \frac{q\left(\predictor_i \mid \predictor^\star_i\right)}{1}
            \right\rbrace ;
            \]
        }
        \tcp{Accept/reject between-model move}
        With probability $\alpha$, set $(\predictors, \indicators) \leftarrow (\predictors^\star, \indicators^\star)$\;
    }
    Set $(\predictors^{(t+1)}, \indicators^{(t+1)}) \leftarrow (\predictors, \indicators)$\;

    \tcp{Within–model update (fixed $\indicators$)}
    Update $\sigma^2$ and coefficients $\predictors$ for included predictors in separate Gibbs steps\;
}
\end{algorithm}

\subsection{Comparing RJMCMC and MoMS for Variable Selection}
Although reversible jump Markov chain Monte Carlo (RJMCMC) and mixtures of mutually singular (MoMS) distributions are fundamentally different approaches to variable selection, the two methods can be made to coincide at the level of their acceptance probabilities. Nevertheless, they remain conceptually distinct, with consequences for the design of proposal distributions and computational implementation.
% Nevertheless, they remain conceptually distinct, with consequences for posterior inference. 

% {\sc NOTE:  I do think this distinction is not as fundamentally different for variable selection where implicitly there is an equivalence to writing a model as $\bm{X} \beta$ and partitioning $\beta = (\beta_\gamma, \beta_{(-\gamma)}$ where $\beta_{(-\gamma)} \equiv \mathbf{0}$ and $\beta$ is $p$ dimensional so that $\mathbf{X} \beta = \bm{X}_\gamma \beta_\gamma$ so that if using the same priors the two lead to exactly the same posterior distribution.  The distinction is really in the proposal distribution used for the coefficients in the new model.}
% Don: Not sure if I fully addressed your comment, now. I think 'posterior inference' was a bit ambiguous here in the sense that it meant computation of the posterior, and not the actual posterior distriubtion. I've now reworded this so there is no ambiguity about that part.

In RJMCMC, moves between models are constructed by augmenting the current state with auxiliary variables and applying a transformation that matches the parameter dimensions of the current and proposed models. While such transformations are used to facilitate efficient transitions, they are not required for detailed balance. In particular, if the transformation $g$ is the identity map, then adding predictor $i$ sets
\[
(\predictors^\star_{\setminus i}, \predictors_i^\star) = (\predictors_{\setminus i}, u).
\]
The auxiliary variable $u$ is the proposed value of the new coefficient. Under this construction, the auxiliary variable coincides with the coefficient proposal, which simplifies the design of the proposal distribution by eliminating the need to account for transformation.

The MoMS procedure outlined in Algorithm~\ref{alg:MoMS} does not involve such a transformation and instead proposes coefficient updates directly using an adaptive random-walk Metropolis step. This generic approach applies broadly and avoids the need for dimension-matching constructions. These design choices have practical implications for sampling efficiency, which we examine in the numerical comparison below. Forster’s procedure exploits model-specific structure through carefully chosen auxiliary-variable proposals. This highlights that efficiency gains in RJMCMC arise from the joint choice of the transformation $g$ and the auxiliary-variable distribution, rather than from the reversible-jump framework per se.

The distinction between coefficient proposals in MoMS and auxiliary-variable proposals in RJMCMC is one of choice rather than necessity. 
% based on feedback
As shown by \citet{DellaportasEtAl_2002} and \citet{godsill_relationship_2001}, if the RJMCMC proposal for new parameters is their exact posterior distribution under the candidate model, the acceptance ratio simplifies to a ratio of marginal likelihoods, effectively recovering algorithms like $MC^3$ \citep{MadiganYork_1995} and recasting the pseudo-prior approach of \citet{CarlinChib_1995} as a special case. From a MoMS perspective, this corresponds to a Gibbs step \citep[see also Example~4 in][]{GottardoRaftery_2008}. The full conditional is again a spike and slab distribution of the form
\begin{align*}
p(\predictor_i \mid y, \bm{X}, \predictor_{-i}) 
=\, &p(\indicator_i = 0\mid y, \bm{X}, \predictor_{-i})
\,\mathbf{1}_{\{0\}}(\predictor_i)\\
&+ p(\indicator_i = 1\mid y, \bm{X}, \predictor_{-i})\,p(\predictor_i\mid y, \bm{X})\,\mathbf{1}_{\mathbb{R}\setminus\{0\}}(\predictor_i),
% I omitted the \mathbf{1} because the equation got too long
\end{align*}
Here, the conditional inclusion probabilities $p(\predictor_i = 0\mid y, \bm{X}, \predictor_{-i})$ and $p(\predictor_i = 1\mid y, \bm{X}, \predictor_{-i})$ can be obtained from the same marginal likelihoods used in methods discussed above.
This illustrates how the MoMS framework could be extended to use more tailored proposal distributions.
% In principle, the MoMS framework could be extended to use more tailored proposal distributions, such as full-conditional proposals \citep{GottardoRaftery_2008} or other model-specific constructions. 
Conversely, RJMCMC can be implemented with simple identity transformations. If both methods use the same proposal density $q$ and the identity map for $g$, then their Metropolis–Hastings acceptance probabilities coincide exactly.

Despite this alignment, MoMS and RJMCMC are not the same algorithm. The key difference lies in how excluded predictors are represented. In RJMCMC, predictors that are not included in the current model are absent from the state space and have no associated parameter values. In MoMS, all coefficients are always present, with excluded predictors represented by coefficients fixed at zero. This distinction affects how the output of the Markov chain can be analyzed. In RJMCMC, marginal chains for individual coefficients may contain long stretches of missing values, so inference is typically restricted to conditional distributions given inclusion. In contrast, MoMS yields well-defined marginal chains for all coefficients. Although coefficients excluded under RJMCMC are often coded as zero, this convention effectively replaces absence from the state space with a point mass at zero and thereby yields a representation closer to MoMS than to the native RJMCMC representation.  %hmm that interpretation predates MOMS. % Don: that may be true, but that doesn't mean this interpretation doesn't fit moms better?

\subsubsection{Numerical Comparison on the Diabetes Dataset}

We illustrate the numerical behavior of enumeration, RJMCMC, and the method of MoMS distributions using the diabetes dataset with $p=10$ predictors. For this problem, full enumeration of all $2^{10}=1024$ models is feasible and provides exact posterior model probabilities, which we use as a benchmark for assessing the sampling methods.

Table~\ref{tb:running_all_methods} reports posterior means and posterior inclusion probabilities obtained by enumeration, RJMCMC, and MoMS. Across all predictors, the two sampling methods closely match the enumeration results, with discrepancies limited to the second decimal place. In particular, all three approaches identify the same predictors as strongly supported (e.g., \textit{BMI} and \textit{BP}), weakly supported (e.g., \textit{AGE} and \textit{S6}), or exhibiting intermediate evidence for inclusion (e.g., \textit{S1}--\textit{S4}). These results confirm that both RJMCMC and MoMS recover the correct posterior distribution for this problem.

%\begin{table}[!ht]
%\centering
%\caption{Posterior means and posterior inclusion probabilities for the running example, obtained by enumeration, RJMCMC, and the method of MoMS distributions.}
%\label{tb:running_all_methods}
%\include{tables/combinedTable.tex}
%\tablenote{The first column denotes the predictor corresponding to the regression coefficient. $\bm{\beta}_{\mathrm{BMA}}$ denotes the model-averaged estimate, $p(\bm{\beta}\neq 0 \mid \mathrm{data})$ the posterior inclusion probability, and Std.\ the posterior standard deviation.}
%\end{table}

\begin{landscape}
\begin{table}[!ht]
\centering
\caption{Exact (enumeration) and Monte Carlo (RJMCMC and MoMS) estimates of regression coefficients and posterior inclusion probabilities for the diabetes dataset}
\label{tb:running_all_methods}
% Maarten heeft kolom namen ge-edit
% Maarten heeft RJ en MoMS omgedraait (volgorde paper)
\begin{tabular}{rrrrrrrrrrrr}
  \toprule
   & \multicolumn{3}{@{}c@{}}{\textbf{$p(\indicator = 1\mid \bm{y}, \bm{X})$}} &  & \multicolumn{3}{@{}c@{}}{$\mathbb{E}(\predictors \mid \bm{y},\bm{X})$} &  & \multicolumn{3}{@{}c@{}}{$\text{SD}(\predictors \mid \bm{y},\bm{X})$} \\
  \cmidrule{2-4}\cmidrule{6-8}\cmidrule{10-12}
   & {Exact} & {RJMCMC} & {MoMS} &  & {Exact} & {RJMCMC} & {MoMS} &  & {Exact} & {RJMCMC} & {MoMS} \\
  \midrule
  AGE & 0.079 & 0.079 & 0.079 &  & -0.001 & -0.001 & -0.001 &  & 0.061 & 0.061 & 0.061 \\
  SEX & 0.987 & 0.987 & 0.987 &  & -21.399 & -21.401 & -21.402 &  & 6.234 & 6.231 & 6.231 \\
  BMI & 1.000 & 1.000 & 1.000 &  & 5.699 & 5.699 & 5.700 &  & 0.712 & 0.713 & 0.713 \\
  BP & 1.000 & 1.000 & 1.000 &  & 1.115 & 1.115 & 1.115 &  & 0.219 & 0.218 & 0.218 \\
  S1 & 0.661 & 0.660 & 0.662 &  & -0.448 & -0.447 & -0.448 &  & 0.464 & 0.464 & 0.464 \\
  S2 & 0.453 & 0.453 & 0.453 &  & 0.260 & 0.260 & 0.261 &  & 0.432 & 0.432 & 0.432 \\
  S3 & 0.515 & 0.516 & 0.514 &  & -0.484 & -0.485 & -0.483 &  & 0.548 & 0.549 & 0.549 \\
  S4 & 0.257 & 0.257 & 0.257 &  & 1.830 & 1.827 & 1.834 &  & 4.324 & 4.320 & 4.324 \\
  S5 & 1.000 & 1.000 & 1.000 &  & 55.362 & 55.340 & 55.379 &  & 14.173 & 14.170 & 14.176 \\
  S6 & 0.125 & 0.125 & 0.125 &  & 0.035 & 0.035 & 0.034 &  & 0.133 & 0.132 & 0.132 \\
  \bottomrule
\end{tabular}
\tablenote{The first column lists the predictors. The left block reports
posterior inclusion probabilities $p(\indicator = 1 \mid \bm{y}, \bm{X})$, the middle block reports model-averaged posterior means $\mathbb{E}(\predictors \mid \bm{y}, \bm{X})$, and the right block reports model-averaged posterior standard deviations $\text{SD}(\predictors \mid \bm{y}, \bm{X})$. Within each block, results are shown for enumeration (Exact), RJMCMC, and MoMS.}
\end{table}
\end{landscape}

To compare computational efficiency, we examine the effective sample size (ESS) of the inclusion indicators. These indicators determine the posterior inclusion probabilities that form the primary inferential target in Bayesian variable selection.
Table~\ref{tb:running_ess} reports the ESS per iteration and per second for the indicator variables. 
%{\sc [AER March 13: I'd suggest removing the three decimal places in the ESS/second columns, so that just the integer value is shown.]}
% Don: tables/essTableGammaRounded.tex contains the rounded values.
% I've left the unrounded ones in the first example because there all the ESS/second are below 1.
\begin{table}[!ht]
\centering
\caption{Effective sample size (ESS) of the inclusion indicators for RJMCMC and MoMS. 
The table reports ESS per iteration and ESS per second for each predictor.}
\label{tb:running_ess}
\begin{tabular}{rrrrrr}
  \toprule
   & \multicolumn{2}{@{}c@{}}{{ESS / iter}} &  & \multicolumn{2}{@{}c@{}}{{ESS / second}} \\
  \cmidrule{2-3}\cmidrule{5-6}
  $\predictor$ & {MoMS} & {RJMCMC} &  & {MoMS} & {RJMCMC} \\
  \midrule
  AGE & 1.317 & 1.308 &  & 47,342 & 7,075 \\
  SEX & 0.315 & 1.124 &  & 11,303 & 6,080 \\
  BMI & . & . &  & . & . \\
  BP & 0.370 & 1.111 &  & 13,311 & 6,012 \\
  S1 & 0.038 & 0.325 &  & 1,357 & 1,759 \\
  S2 & 0.040 & 0.375 &  & 1,423 & 2,029 \\
  S3 & 0.026 & 0.171 &  & 927 & 924 \\
  S4 & 0.078 & 0.430 &  & 2,790 & 2,329 \\
  S5 & 0.121 & 1.111 &  & 4,359 & 6,012 \\
  S6 & 1.189 & 1.477 &  & 42,722 & 7,992 \\
  \bottomrule
\end{tabular}
\tablenote{The ESS for the BMI indicator is not reported because the effect was included in all post-warmup iterations, resulting in no variation in the indicator sequence.}
\end{table}

For RJMCMC, coefficient values are defined only when the corresponding effect is included in the model. Consequently, a continuous chain for the regression coefficients does not exist across all iterations. In contrast, under the MoMS formulation coefficients remain part of the state space and are set to zero when excluded. To ensure a comparable efficiency measure across methods, we therefore compute ESS based on the indicator sequences rather than the coefficients.

The ESS for the indicators is derived from the binary Markov chain representation described in Appendix~\ref{App::ess}. Table~\ref{tb:running_ess} shows that RJMCMC achieves somewhat higher ESS per iteration for the indicators, whereas MoMS achieves substantially higher ESS per second. This pattern reflects a trade-off: more elaborate proposal mechanisms can improve mixing per iteration but may also increase computational cost. In the present implementation, RJMCMC uses carefully constructed dimension-matching proposals, whereas MoMS relies on comparatively inexpensive adaptive Metropolis updates. As a result, MoMS iterations are cheaper, leading to competitive efficiency when measured in wall-clock time.

\section{Case Study: Variable Selection in Mixed Logistic Regression}

We illustrate the method using the data of  \citet{mak2024data}, who studied the effect of sleep on memory recall. Their experiment used a 2 x 2 between-participants design crossing Immediacy of recall (immediate vs. 12-hour delay) with Time of day (AM vs. PM). Participants learned lists of words and were later tested using separate lists containing both studied words and lures.

The study followed the Deese-Roediger-McDermott (DRM) paradigm \citep{deese1959prediction, roediger1995creating}, in which participants study lists of semantically related words (e.g., medicine, hospital, physician) but not the associated ``critical lure'' (e.g., doctor), which is often falsely recalled.

The analyzed dataset contains $488$ participants, each studying $20$ lists comprising $160$ items in total. Each list contains one critical lure, resulting in 20 lure words.

We model the probability of recall using a mixed-effects logistic regression:
\begin{align*}
    \mathrm{logit}(P(\mathrm{Recall}_{ij} = 1))
    &= \beta_0 + \beta_1 \text{Lure}_{j} + \beta_2 \mathrm{Immediacy}_{i} + \beta_3 \mathrm{Time}_{i} \\
    & + \beta_4 \mathrm{\#Intrusions} \\
    &+ \beta_5 (\mathrm{Immediacy}_{i} \times \mathrm{Time}_{i}) + \beta_6(\mathrm{Lure}_{j}\times \mathrm{\#Intrusions}) \\
    &+ \beta_7 (\mathrm{Lure}_{j} \times \mathrm{Immediacy}_{i} \times \mathrm{Time}_{i}) \\
    &+  u_{i} + w_{j}
\end{align*}
Here, $u_{i}$ and $w_{j}$ denote random intercepts for subjects $i$ and words $j$, respectively.

% {\sc MAC: rest of model specification?   Priors on fixed effects and random effects?}
% Done right above "In total, we obtained..." two paragraphs down.

The intercept $\beta_0$ and the fixed effects $\beta_1$ through $\beta_5$ correspond directly to the preregistered hypotheses of \citet{mak2023registered}. Evidence that $\beta_1 \neq 0$ reflects the classic DRM false memory effect, indicating that lure words differ in retrieval rate from studied words. The effects of Immediacy ($\beta_2$), Time of day ($\beta_3$), and their interaction ($\beta_5$) assess whether recall differs across delay conditions and whether a 12-hour interval that includes sleep differs from a daytime wake interval. The coefficient $\beta_4$ captures the association between recall and the number of intrusions.

Bayesian variable selection was performed over all fixed effects, excluding the intercept, using the MoMS construction described in Section~\ref{Sec:MoMS}. Thus, main effects and interaction terms were jointly considered for inclusion. To facilitate posterior computation, P{\'o}lya--Gamma data augmentation \citep{polson2013bayesian} was used, which yields Gaussian updates for the regression coefficients.
Standard normal prior distributions were assigned to all fixed effects parameters. 
For the random effects, zero-mean normal prior distributions were specified, with the variance drawn from a positive Cauchy distribution with location $0$ and scale $2.5$ (following, \citealp{gelmanPriorDistributionsVariance}, but see \citealp{gelmanPriorCanOften2017} for other recommendations).
In total, $50,000$ posterior samples were obtained.
The $\hat{R}$ was smaller than $1.005$ for all parameters.
% changes after inverse-gamma to half-cauchy: 
% - [x] updated the no. samples from 10,000 to 50,000 (since I had to rerun it anyway) 
% - [x] rhat if needed (not needed)
% - [x] table 3 (minor changes in decimals)
% - the random effect standard deviations went from
% $\hat{\sigma}_u = 0.747\ 95\% \mathrm{CRI}: [0.694, 0.804]$ and $\hat{\sigma}_v = 0.632\ 95\% \mathrm{CRI}: [0.564, 0.708]$
% to 
% $\hat{\sigma}_u = 0.747\ 95\% \mathrm{CRI}: [0.694, 0.804]$ and $\hat{\sigma}_v = 0.632\ 95\% \mathrm{CRI}: [0.565, 0.709]$
% (for $\hat{\sigma}_v$ the credible interval increased by 0.001).

Table~\ref{tb:logisticResultsFixedEff} reports posterior inclusion probabilities and parameter summaries for the fixed effects. The posterior means of the random-effect standard deviations are $\hat{\sigma}_u = 0.747\ 95\% \mathrm{CRI}: [0.694, 0.804]$ and $\hat{\sigma}_v = 0.632\ 95\% \mathrm{CRI}: [0.565, 0.709]$, closely matching the corresponding frequentist estimates.

\begin{table}[!ht]
    \centering
    \caption{Fixed Effects in the Mixed Logistic Regression Model}
\begin{tabular}{rrrrr}
  \toprule
  Parameter & $p(\gamma\mid\bm{y},\bm{X})$ & $\mathbb{E}(\bm{\beta}\mid\bm{y},\bm{X})$ & $\mathrm{SD}(\bm{\beta}\mid\bm{y},\bm{X})$ & $\mathrm{ESS}/\mathrm{second}$ \\
  \midrule
  Intercept & . & -2.0907 & 0.0833 & . \\
  Lure & 0.1495 & -0.0119 & 0.0675 & 1.8151 \\
  Delay: delay & 0.9914 & -0.4559 & 0.1200 & 0.0522 \\
  Time: PM & 0.2236 & 0.0347 & 0.0851 & 0.3971 \\
  \#intrusions & 0.0034 & -0.0000 & 0.0029 & 1.4319 \\
  Delay: delay $\times$ Time: PM & 0.8041 & -0.2526 & 0.1792 & 0.2418 \\
  Lure $\times$ \#intrusions & 0.0045 & -0.0000 & 0.0030 & 2.6214 \\
  Lure $\times$ Delay: delay & 1.0000 & 0.4813 & 0.0884 & . \\
  Lure $\times$ Time: PM & 0.0974 & 0.0022 & 0.0304 & 4.1176 \\
  Lure $\times$ Delay: delay $\times$ Time: PM & 0.4446 & 0.0923 & 0.1241 & 1.0800 \\
  \bottomrule
\end{tabular}
    \tablenote{The inclusion probability for the intercept is omitted because it was included in all models. Reference categories are omitted for the interactions effects. For all parameters, including random effects, $\rhat < 1.005$. The final columns report efficiency as effective sample size per second. As in Table~\ref{tb:running_ess}, the ESS for the intercept is not reported because it was included in all iterations.} % why is the spacing so weird?
    \label{tb:logisticResultsFixedEff}
\end{table}

\begin{figure}[!ht]
    \caption{Sample means (dots) and box plots for the Lure $\times$ Delay contrast.}
    \begin{center}
    \includegraphics[width=.5\linewidth]{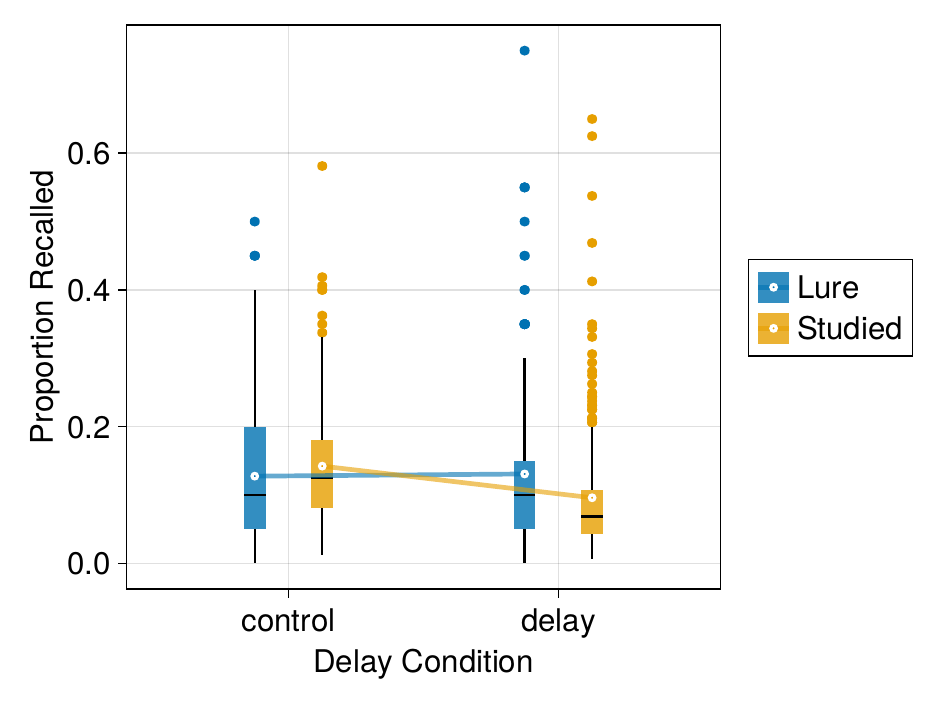}
    \end{center}
     \figurenote{The pattern observed in the posterior estimates is also visible in both the means and medians of the observed data.}
    \label{fig:interactionLureDelay}
\end{figure}

The reference condition (intercept) corresponds to recall of non-lure words in the Immediate-AM group. Relative to this baseline, there is evidence for a negative effect of Delay, indicating reduced recall after a 12-hour interval. There is little evidence for a main effect of Time of day. However, the Immediacy x Time interaction is supported, indicating that the effect of delay depends on time of testing. Specifically, recall following a 12-hour interval that includes overnight sleep exceeds recall following a daytime wake interval. Figure~\ref{fig:interactionLureDelay} visualizes the interaction between Immediacy and Lure in the raw data.

\section{Case Study: Item-Factor Selection in the Multidimensional Generalized Partial Credit Model}
We now consider item–factor selection in a multidimensional item response model. Specifically, we apply MoMS to the multidimensional generalized partial credit model (MGPCM), where sparsity in the factor loadings determines the dimensional structure.

\subsection{The Multidimensional Generalized Partial Credit Model}
The MGPCM extends the generalized partial credit model \citep{muraki1992generalized} to multiple latent dimensions. Suppose we observe scores from $P$ participants on $I$ items, where the response of participant $p$ on item $i$ takes  values $y_{pi} \in \{0, 1, \dots, K_{i}-1\}$. Let $\bm{\theta}_p \in \mathbb{R}^D$ denote the vector of latent traits for participant $p$, and let $\bm{\alpha}_i \in \mathbb{R}^D$ denote the factor loadings of item $i$. The probability of observing a response in category $k$ is 
\begin{align*}
    \prob{y_{pi} = k\mid \theta_p, \alpha_i, \bm{\beta}_i}
    =
    \frac{\exp\left(
    \sum_{r=0}^k \bm{\alpha}_i^T\bm{\theta_p} - \beta_{ri}
    \right)}{
    \sum_{v=0}^{K_i}
    \exp\left(
    \sum_{r=0}^v \bm{\alpha}_i^T\bm{\theta_p} - \beta_{ri}
    \right)}.
\end{align*}
Letting $\beta^\ast_{jk} = \sum_{r=0}^k \beta_{jr}$ we may simplify the model definition to
\begin{align*}
    \prob{y_{pi} = k\mid \theta_p, \alpha_i, \bm{\beta}^\ast_{i}}
    =
    \frac{\exp\left(
    k\bm{\alpha}_i^T\bm{\theta_p} - \beta^\ast_{ik}
    \right)}{
    \sum_{v=0}^{K_i}
    \exp\left(
    v\bm{\alpha}_i^T\bm{\theta_p} - \beta^\ast_{iv}
    \right)}
\end{align*}
Here, $\beta^\ast_{ik}$ are threshold parameters, with $\beta^\ast_{i0} = 0$ for identification.

The model is invariant under orthogonal rotations of the latent space: For any rotation matrix $\mathbf{R}$ with $\mathbf{R}^{\sf T}\mathbf{R} = \mathbf{I}$, the transformation $\bm{\theta}_p \mapsto \mathbf{R}\bm{\theta}_p$ and $\bm{\alpha}_i \mapsto \mathbf{R}\bm{\alpha}_i$ leaves the likelihood unchanged. To achieve identification, we impose a lower-triangular constraint on the loading matrix by requiring that $\alpha_{ii} > 0$ and $\alpha_{ij} = 0$ for $i < j$, following \citep{MavridisNtzoufras_2014}.

In multidimensional IRT models, the factor loadings $\bm{\alpha}$ determine the dimensional structure of the model.
Sparsity in $\bm{\alpha}$ indicates which items load on which latent dimensions and therefore forms the basis for structural interpretation. In the following, variable selection is performed over the free factor loadings using MoMS.
As prior distributions, we use $p(\theta_{pd}) = \dnorm{0,1}$, $p(\beta^\ast_{ik}) = \dnorm{0,1}$ and for the slab $p(\alpha_{ij}) = \dnorm{0,1}$ if $i>j$. For the diagonal of the factor loadings, we use a moderately informed prior, $p(\alpha_{ii}) = \dinvgamma{\nicefrac{31}{3}, \nicefrac{140}{9}}$, which has a prior mean of $\nicefrac{5}{3}$ and variance of $\nicefrac{1}{3}$. This parameterization is chosen specifically to push those factor loadings away from zero. If these elements approach zero, the model loses its structural identification and the aforementioned rotational invariance returns.

% {\sc MAC: rest of model specification?   Priors ?}
% Done, see above

\subsection{Stigmatizing Attitudes among Health Care Workers}
We analyze a dataset on stigmatizing attitudes of health care workers toward people with mental distress.
\citet{daguman2025evaluation} examined the psychometric properties of the 15-item
Opening Minds Scale for Health Care Providers \citep[developed by][]{kassam2012development}.
The full dataset comprises $286$ participants; after excluding 7 participants with missing values, $279$ remain for analysis.

We fit a three-dimensional MGPCM to the data and perform variable selection over all free factor loadings using the MoMS construction described above. The lower-triangular identification constraints on the loading matrix are maintained throughout sampling. The threshold parameters and latent traits are estimated in all models. Posterior inclusion probabilities and model-averaged loading estimates are obtained from the Markov chain output. We discarded the first 20,000 posterior samples as warmup and based the results on the the 200,000 samples drawn afterward. The $\hat{R}$ was below 1.005 for both the full model and the MoMS model.

\begin{figure}[!ht]
    \caption{Shrinkage of Factor Loadings under Model Averaging}
    \begin{center}
    \includegraphics[width=\linewidth]{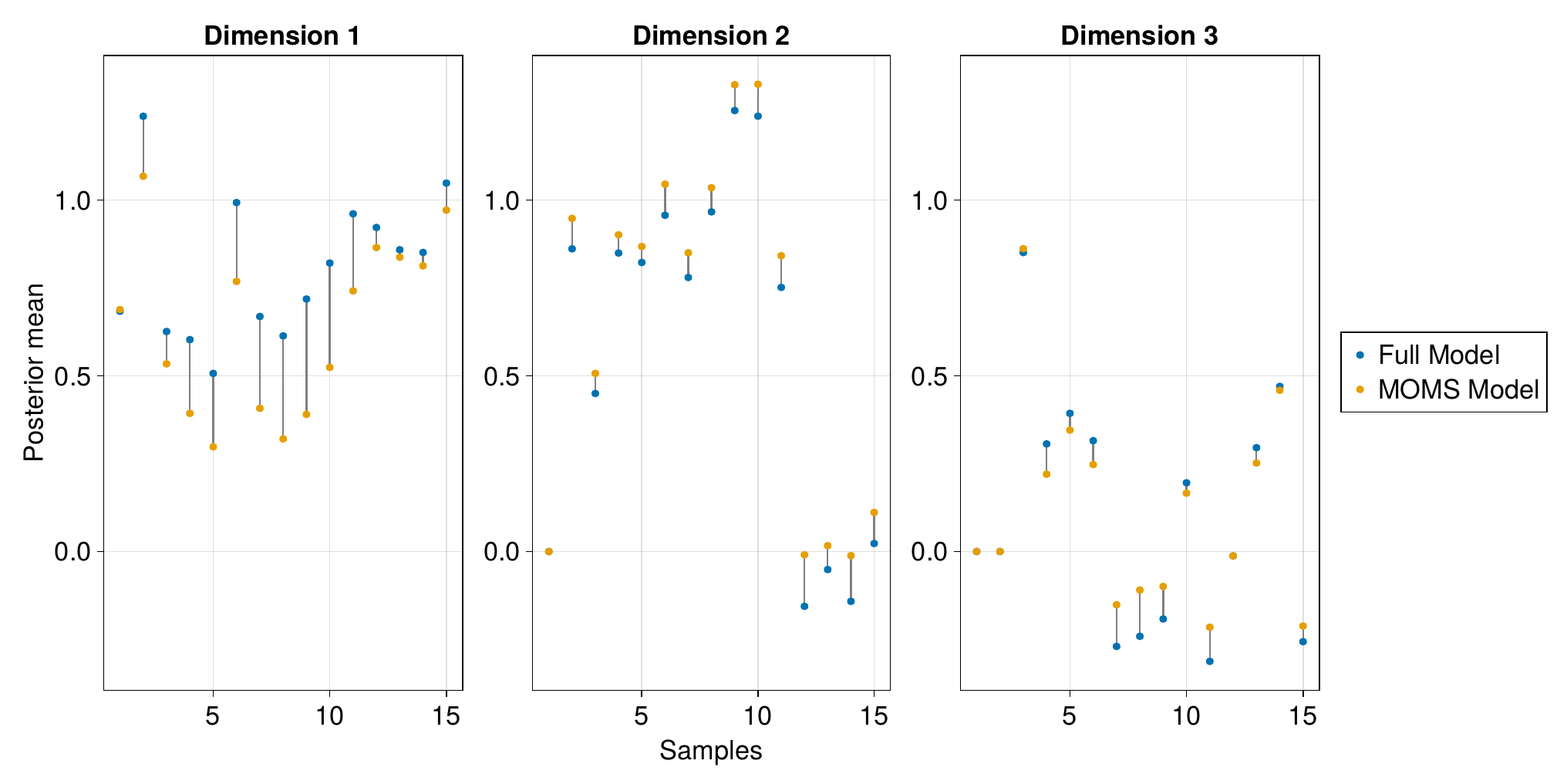}
    \end{center}
    \figurenote{Blue points denote factor loadings under the full model; orange points denote Bayesian model-averaged loadings.}
    \label{fig:gpcm_shrinkage}
\end{figure}

Figure~\ref{fig:gpcm_shrinkage} displays the effect of variable selection on the factor loadings. Loadings on dimensions 1 and 3 are predominantly attenuated under model averaging, reflecting shrinkage induced by selection. In contrast, several loadings on dimension 2 increase relative to the full-model estimates. This pattern is consistent with posterior dependence among loadings: as some loadings are shrunk toward zero, others adjust to maintain model fit. Figure~\ref{fig:gpcm_norm_shrinkage} shows that the Euclidean norm of the loading vector for each item decreases under model averaging, indicating that the overall magnitude of item influence is reduced.

\begin{figure}[!ht]
    \caption{Factor Loadings under the Median Probability Model}
    \begin{center}
    \includegraphics[width=\linewidth]{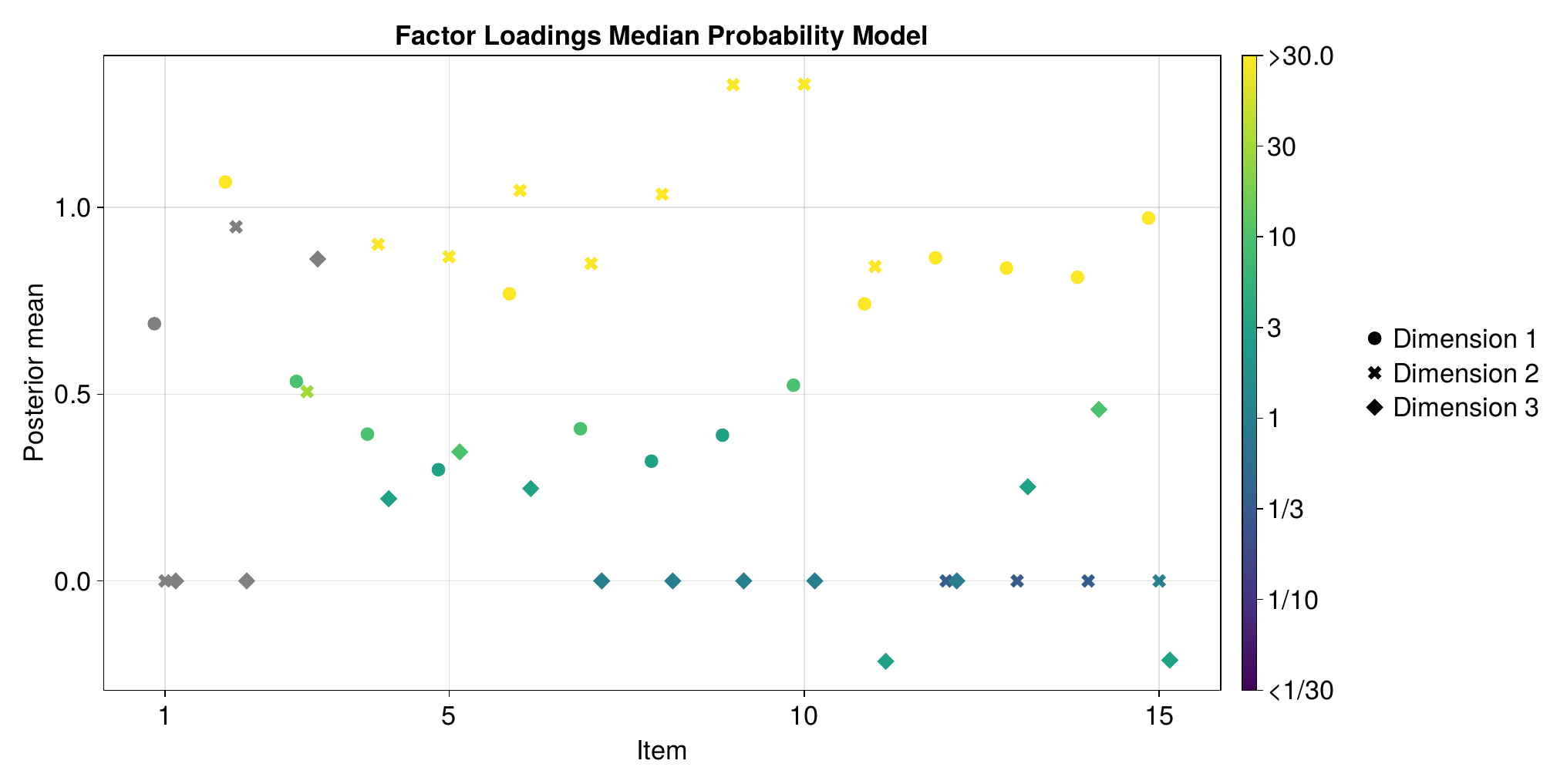}
    \end{center}
    \figurenote{Estimates under the median probability model, retaining loadings with posterior inclusion probability exceeding $\nicefrac{1}{2}$. Excluded loadings are set to zero. Color indicates the inclusion Bayes factor and shape indicates the dimension. Grey points correspond to loadings fixed by identification constraints.}
    \label{fig:gpcm_mpm}
\end{figure}

Model-averaged estimates typically remain dense. To obtain a sparse representation, we therefore consider the median probability model \citep{BarbieriBerger_2004, ghosh2015bayesian}, retaining loadings with posterior inclusion probability exceeding $0.5$. 
% rationale for why sparse solution is necessary?  Wouldn't model averaging be a better option here?   What is the objective for using the median probability model, which can be sub-optimal with correlated predictors?   For insight?
% Don: Yes, the goal of many of these scale validation studies is to find evidence that the items nicely load on their intended factors. Here that is clearly not the case, although the original study did assume the structure where OA1-OA18 loaded on Att, OD4-OD10 load on DHS, and OS3-OS19 load on ScD, and then used a variety of goodness of fit measures.
Figure~\ref{fig:gpcm_mpm} presents the resulting structure along with the corresponding inclusion Bayes factors.

For loading $j$, the inclusion Bayes factor compares posterior inclusion odds to prior inclusion odds. The prior and posterior inclusion probabilities are
\begin{align}
p(\indicator_j = 1) &= \sum_{\indicators : \indicator_j = 1} p(\indicators), \\
p(\indicator_j = 1 \mid \bm{X}) &= \sum_{\indicators : \indicator_j = 1} p(\indicators \mid \bm{X}),
\end{align}
which gives
\begin{equation}
\mathrm{BF}_{\mathrm{incl},j}
=
\frac{p(\indicator_j = 1 \mid \bm{X}) / p(\indicator_j = 0 \mid \bm{X})}
     {p(\indicator_j = 1) / p(\indicator_j = 0)}.
\end{equation}

\begin{figure}[!ht]
    \caption{Empirical Factor Structure under the Median Probability Model}
    \begin{center}
    \includegraphics[width=\linewidth]{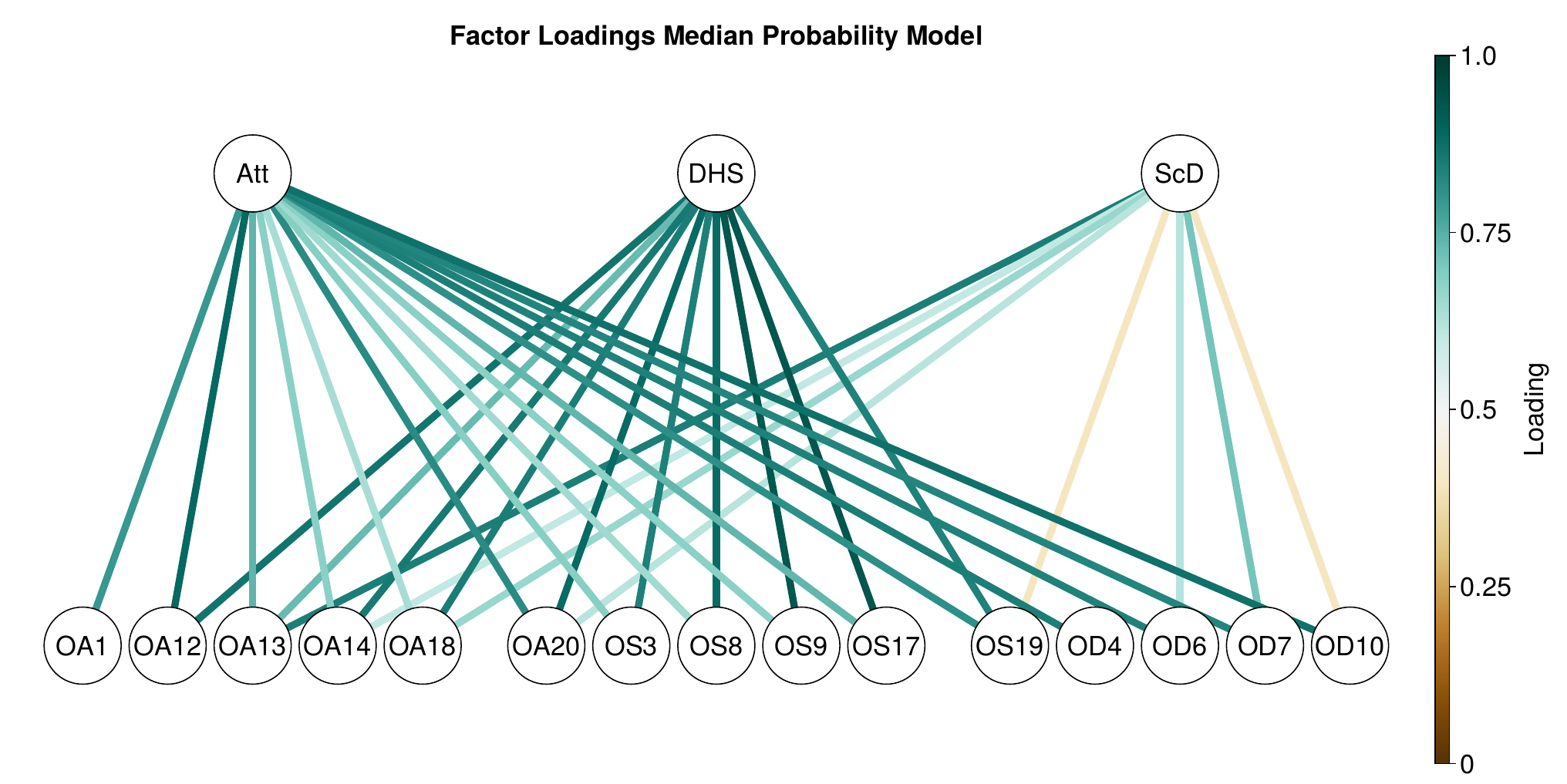}
    \end{center}
    \figurenote{Item loadings on the three latent dimensions. Abbreviations: Attitude (\emph{Att}), Disclosure and Help-seeking (\emph{DHS}), and Social Distance (\emph{ScD}).}
    \label{fig:gpcm_factor_structure}
\end{figure}

Figure~\ref{fig:gpcm_factor_structure} indicates that items load more broadly across dimensions than suggested by the theoretical structure, yielding a denser empirical factor pattern.

\section{Discussion}

In this paper, we presented mixtures of mutually singular (MoMS) distributions as a transparent framework for Bayesian variable selection. Our aim was to show that, for nested variable-selection problems, MoMS provides a coherent representation of model uncertainty while targeting the same posterior quantities of interest as reversible-jump Markov Chain Monte Carlo (RJMCMC). We further showed that, under appropriate constructions, MoMS and RJMCMC can yield identical Metropolis--Hastings acceptance probabilities and, in those cases, reduce to essentially the same algorithm. In the examples considered here, this equivalence was accompanied by comparable computational performance while preserving the interpretability of spike-and-slab variable selection.

The main implication is not that MoMS should replace RJMCMC in every transdimensional setting, 
% Indeed there are cases were they cannont!  Look at my LARK models in AOA or Valencia 7 volume - the "model space" is infinite dimensional
% Don: I agree, but isn't this clear from how we phrase this already? I'm not sure how you'd like to address this, we do not do anything infinite dimensional here or anywhere else?
but that it is especially natural for Bayesian variable selection in nested model classes. In that setting, the framework makes the model space explicit through mutually singular subspaces and allows posterior inclusion and model-averaged inference to be formulated directly. The remainder of this discussion considers three broader points: how the framework extends to multiple-effect updates, how the same logic can be used to define more general exclusive-subspace models, and where its advantages become less clear.

\subsection{Multiple Updates for Variable Selection}

We first consider an extension that remains within the nested variable-selection setting studied throughout this paper: updating multiple effects jointly rather than one effect at a time. This extension becomes useful when posterior dependence makes one-at-a-time updates inefficient, so that coordinated moves may improve exploration of the model space. For two effects $i$ and $j$, the proposal distribution can be factorized as
\[
p(\indicator_i, \predictor_i, \indicator_j, \predictor_j \mid \cdots)
=
p(\indicator_i, \predictor_i \mid \indicator_j, \predictor_j, \cdots)
\, p(\indicator_j, \predictor_j \mid \cdots).
\]
This factorization shows that block proposals can be constructed in a principled way, either by chaining conditional proposals or by proposing jointly on $(\indicator_i,\predictor_i,\indicator_j,\predictor_j)$. From a Metropolis--Hastings perspective, this extension does not introduce a new conceptual difficulty: proposal normalizing constants enter symmetrically in the acceptance ratio and cancel in the usual way. Moving from single-effect to multiple-effect updates is therefore primarily a matter of proposal design and mixing behavior.

In MoMS, this extension is especially transparent because it follows directly from the factorization of the proposal distribution. Comparable block moves can in principle be implemented in RJMCMC as well, although it may be less obvious how best to construct the corresponding between-model proposals \citep[but see][section 7.3.3]{fan2026reversible}. 
% See Dennison et al (JRSSB 1998) for RJMCMC and discussion in Nott and Green (JCGS 2004) for block updates.  The simple swap step is what is used on BAS with integrated marginal likelihoods. Overcomes waiting time problem for leaving a high probability state to a low probability state that would permit moving to another high probability state with inclusion of one of two variables that are highly correlated.  More sophisticated versions can be developed that take advantage of correlation structure of the predictors but this simpler approach is efficient and improves mixing over one-at-a-time updates.
% Don: I've added a reference to the subsection "Multi-Step Proposals" from the Handbook of MCMC (2026 edition), which also discusses this.
This also connects MoMS-based variable selection to the literature on informed proposals in discrete spaces, where locally balanced constructions provide a principled way to design multi-coordinate moves and, in high-dimensional local-update settings, can improve efficiency; when fully informed proposals become costly, block-wise implementations provide a practical compromise \citep[e.g.,][]{zanella2020informed}.

\subsection{Variable-Selection Models Over Exclusive Subspaces}

A broader implication of the MoMS framework is that it can be used not only for add/delete variable selection, but also to define model classes through mutually exclusive regions of the parameter space. On this view, model uncertainty need not be tied only to the presence or absence of an effect. It can also be represented through support restrictions that correspond to substantively distinct hypotheses.

Examples include positive, zero, and negative regions for directional effects; a region of practical equivalence around zero versus its complement, which reformulates the ROPE procedure \citep{KruschkeLiddell_2018} as variable selection; or a finite set of point predictions versus a free continuous alternative. In each case, the parameter space is partitioned into exclusive subspaces, and posterior inference proceeds across those subspaces rather than within a single unconstrained model. These examples are best viewed as sketches, but they illustrate that the same logic underlying variable selection extends to a broader class of model-comparison problems.

In MoMS, such constructions are specified directly through support restrictions. Similar ideas can in principle be implemented in RJMCMC, but the modeling path is often less transparent because dimension-matching transformations and reverse moves must be specified simultaneously for each transition. A conceptual advantage of the MoMS framework is therefore that these exclusive-subspace formulations can be expressed directly at the level of the model space.

\subsection{Non-Nested and Broader Transdimensional Settings}

The advantages of MoMS are clearest in nested variable-selection problems, but they do not extend uniformly beyond that setting. RJMCMC may be more natural when the candidate models are genuinely non-nested and do not admit a practically useful common parameterization.
This includes settings in which the model changes type, not only dimension. In such cases, forcing all models into a single fixed-dimensional MoMS representation may be artificial or computationally inefficient, whereas RJMCMC can work more directly with model-specific parameterizations \citep[e.g., in symbolic regression, see][]{jinBayesianSymbolicRegression2020a}. 
% For infinite-dimensional settings such as Bayes non-parametric regression using stochastic expansions such as Chu, Clyde, Liang (2009 Statistic Sinica), Wolpert, Clyde and Tu (AOS 2011) and House, Clyde, and Wolpert (AOAS 2011) "predictor" in the mean function are selected from continuous dictionaries or over-complete representaiona. While birth and death steps may have MoMS like interpretations, additional steps involving merge and split do not and it is not feasible to represent this as a fixed dimensional problem. 
% Don: I've added (and cited) symbolic regression which I think exemplifies the change in model type even better.
For nested variable-selection classes, MoMS can simplify both implementation and interpretation; in broader transdimensional model spaces, however, RJMCMC may offer a simpler route because between-model moves can be designed in the coordinates natural to each model. Which framework is easier in practice therefore depends on the structure of the model space and on the kinds of transitions the sampler must support.

\subsection{Conclusion}

For Bayesian variable selection in nested model classes, MoMS offers a transparent inferential framework in which model uncertainty is represented directly through mutually singular subspaces, while posterior inclusion and model-averaged quantities remain the main targets. In this setting, MoMS preserves the spike-and-slab interpretation, accommodates both single- and multiple-effect updates, and allows substantive constraints to be encoded explicitly in the model space.

More broadly, our results show that Bayesian variable selection need not require abandoning standard MCMC methodology. When the model space admits a shared parameterization, MoMS represents model uncertainty within a fixed-dimensional framework while retaining the exact spike-and-slab interpretation. In some broader transdimensional settings, however, RJMCMC may remain the more natural choice because between-model moves can be formulated in model-specific coordinates.

The broader point is therefore methodological: when a common parameterization is natural, Bayesian variable selection can be carried out within standard MCMC methodology, without regret.

\small
\paragraph{Acknowledgments.} DvdB and MM were supported by the European Union {(ERC, BAYESIAN P-NETS, \#101040876)}. Views and opinions expressed are however those of the author(s) only and do not necessarily reflect those of the European Union or the European Research Council. Neither the European Union nor the granting authority can be held responsible for them.
\normalsize

\bibliography{Marsman.bib, references.bib}

\appendix

%The next two lines we need to add to manually to every \Section in the appendix.
%We also need to manually change {A} to {B}, etc., to label the \Sections.
\renewcommand{\thesection}{A}
\setcounter{subsection}{0}

\section{The Detailed Balance Condition for Markov Chains}\label{App::detailed_balance}

The detailed balance condition expresses equilibrium for a Markov chain.
At equilibrium, transitions of the chain do not change the target distribution:
the probability of moving from one state to another is exactly balanced by the
probability of moving back.

Let $\pi(x)$ denote the target distribution (for example, a posterior density),
and let $P(x \to x^\star)$ denote the one-step transition probability from the
current state $x$ to a new state $x^\star$.
The detailed balance condition is the pointwise equality
\begin{equation}
\pi(x)\, P(x \to x^\star)
=
\pi(x^\star)\, P(x^\star \to x),
\label{eq:DB_pointwise}
\end{equation}
which is required to hold for all states $x$ and $x^\star$ that the chain can move between.

Each side of Eq.~\eqref{eq:DB_pointwise} has a direct probabilistic interpretation.
When the state space is continuous, $\pi(x)$ is a probability density and
$P(x \to x^\star)$ is a transition density, so their product is a joint density
for the event that the chain is at $x$ and moves to a state near $x^\star$ in one
iteration.
Integrating this joint density over regions of the state space yields the
corresponding transition probabilities.
If Eq.~\eqref{eq:DB_pointwise} holds for all pairs of states, then the target
distribution $\pi$ is invariant under the Markov chain.

Many Markov chain algorithms proceed by attempting a move from the current state
$x$ to a candidate state $x^\star$, and then deciding whether to accept this move.
Suppose that, from a current state $x$, a candidate state $x^\star$ is proposed
according to a proposal density $q(x^\star \mid x)$, and that the proposed move is
accepted with probability $\alpha(x \to x^\star)$; if it is rejected, the chain
remains at the current state $x$.

To connect detailed balance to the mechanics of such algorithms, it is helpful to
consider probabilities of moving between regions of the state space.
Let $A$ and $B$ be two subsets of the parameter space.
At equilibrium, the probability of moving from $A$ to $B$ in one iteration must
equal the probability of moving from $B$ back to $A$, that is,
\begin{equation}
\int_A \pi(x)
\left(
\int_B P(x \to x^\star)\, dx^\star
\right) dx
=
\int_B \pi(x^\star)
\left(
\int_A P(x^\star \to x)\, dx
\right) dx^\star.
\label{eq:DB_regions}
\end{equation}

Under the accept--reject mechanism described above, the one-step transition
probability density can be written as
\begin{equation}
P(x \to x^\star)
=
q(x^\star \mid x)\,\alpha(x \to x^\star)
+
\left(
1 - \int q(z \mid x)\,\alpha(x \to z)\,dz
\right)
\mathbf 1_{\{x\}}(x^\star),
\label{eq:P_decomp}
\end{equation}
where the first term corresponds to accepted moves and the second term corresponds
to rejection, which produces a self-transition.

Substituting \eqref{eq:P_decomp} into the left-hand side of
Eq.~\eqref{eq:DB_regions} yields
\begin{align}
&\int_A \pi(x)
\left(
\int_B q(x^\star \mid x)\,\alpha(x \to x^\star)\,dx^\star
\right) dx
\nonumber\\
&\quad+\;
\int_A \pi(x)
\left(
1 - \int q(z \mid x)\,\alpha(x \to z)\,dz
\right)
\mathbf 1_B(x)\,dx.
\label{eq:LHS_split}
\end{align}
An analogous decomposition is obtained for the right-hand side.

The second term in \eqref{eq:LHS_split} contains the indicator $\mathbf 1_B(x)$.\footnote{
In the rejection term of \eqref{eq:P_decomp}, the factor
\(
1-\int q(z\mid x)\,\alpha(x\to z)\,dz
\)
does not depend on $x^\star$.
All dependence on $x^\star$ is contained in the indicator $\mathbf 1_{\{x\}}(x^\star)$.
Consequently,
\[
\int_B
\left(
1-\int q(z\mid x)\,\alpha(x\to z)\,dz
\right)
\mathbf 1_{\{x\}}(x^\star)\,dx^\star
=
\left(
1-\int q(z\mid x)\,\alpha(x\to z)\,dz
\right)\mathbf 1_B(x).
\]
}
Since the outer integral is over $x \in A$, the integrand is nonzero only when
$x \in A \cap B$, and the term can be written as
\[
\int_{A \cap B}
\pi(x)
\left(
1 - \int q(z \mid x)\,\alpha(x \to z)\,dz
\right) dx.
\]
The corresponding rejection term on the right-hand side has the same form.
Hence, the rejection terms on the two sides of Eq.~\eqref{eq:DB_regions} are
identical and cancel exactly.

After cancellation of these terms, Eq.~\eqref{eq:DB_regions} reduces to
\[
\int_A \pi(x)
\left(
\int_B q(x^\star \mid x)\,\alpha(x \to x^\star)\,dx^\star
\right) dx
=
\int_B \pi(x^\star)
\left(
\int_A q(x \mid x^\star)\,\alpha(x^\star \to x)\,dx
\right) dx^\star.
\]
Since this equality must hold for all choices of regions $A$ and $B$ in the state
space, the integrands must agree pointwise.
This yields the condition
\[
\pi(x)\, q(x^\star \mid x)\,\alpha(x \to x^\star)
=
\pi(x^\star)\, q(x \mid x^\star)\,\alpha(x^\star \to x),
\]
which is sufficient for detailed balance.

\renewcommand{\thesection}{B}
\setcounter{subsection}{0}

\section{Solving the Detailed Balance Equation}\label{App::metropolis_hastings}

For the posterior distribution $p(\predictors, \indicators \mid \bm{y}, \bm{X})$ to be invariant, the transition probabilities must satisfy the detailed balance condition
\begin{align}\label{eq:DB_appendix}
&p(\predictors, \indicators \mid \bm{y}, \bm{X})\,
q\!\left(\predictors^\star, \indicators^\star \mid \predictors, \indicators\right)\,
\alpha\!\left((\predictors, \indicators)\!\to\!(\predictors^\star, \indicators^\star)\right)
\\[4pt]
\nonumber
&\qquad =\;
p(\predictors^\star, \indicators^\star \mid \bm{y}, \bm{X})\,
q\!\left(\predictors, \indicators \mid \predictors^\star, \indicators^\star\right)\,
\alpha\!\left((\predictors^\star, \indicators^\star)\!\to\!(\predictors, \indicators)\right).
\end{align}
Here $q(\cdot\mid\cdot)$ denotes the proposal distribution and $\alpha(\cdot)$ the acceptance probability. We will assume that the proposal density $q$ is known and define the acceptance probability $\alpha$ for which detailed balance holds.

Define the ratio of the posterior and transition densities in Eq.~\eqref{eq:DB_appendix}:
\[
\text{R} =
\frac{
p(\predictors^\star, \indicators^\star \mid \bm{y}, \bm{X})\,
q\!\left(\predictors, \indicators \mid \predictors^\star, \indicators^\star\right)
}{
p(\predictors, \indicators \mid \bm{y}, \bm{X})\,
q\!\left(\predictors^\star, \indicators^\star \mid \predictors, \indicators\right)
}.
\]
Eq.~\eqref{eq:DB_appendix} can be written as
\begin{equation}
\alpha\!\left((\predictors, \indicators)\!\to\!(\predictors^\star, \indicators^\star)\right)
=
\text{R}\,
\alpha\!\left((\predictors^\star, \indicators^\star)\!\to\!(\predictors, \indicators)\right)
\quad \Longrightarrow \quad
\alpha_{\text{fwd}} = \text{R}\,\alpha_{\text{rev}}.
\label{eq:alpha_relation}
\end{equation}
To satisfy detailed balance, any pair of acceptance probabilities $\alpha_{\text{fwd}}$ and $\alpha_{\text{rev}}$ that obey this relation will work.

\citet{metropolis1953equation} solved for the acceptance probabilities by setting the larger of the two to one and then solving for the other.
Suppose that $\text{R} \ge 1$. Then $\alpha_{\text{fwd}} \ge \alpha_{\text{rev}}$.
In this scenario, \citet{metropolis1953equation} set $\alpha_{\text{fwd}} = 1$ and $\alpha_{\text{rev}} = 1 / \text{R}$. When $\text{R} < 1$, $\alpha_{\text{fwd}} < \alpha_{\text{rev}}$. In this case, $\alpha_{\text{rev}} = 1$ and $\alpha_{\text{fwd}} = \text{R}$.
Thus, the acceptance probabilities are
\[
\alpha_{\text{fwd}} = \min\!\left(1,\text{R}\right)
\quad \text{and} \quad
\alpha_{\text{rev}} = \min\!\left(1,\tfrac{1}{\text{R}}\right).
\]
This is a solution to Eq.~\eqref{eq:alpha_relation}:
\begin{align*}
\alpha_{\text{fwd}} &= \text{R}\,\alpha_{\text{rev}},\\
\min\!\left(1,\text{R}\right) &= \text{R}\,\min\!\left(1,\tfrac{1}{\text{R}}\right),\\
\min\!\left(1,\text{R}\right) &= \min\!\left(\text{R},1\right).
\end{align*}

There are alternative ways to define acceptance probabilities that satisfy the detailed balance condition.
Instead of setting the larger of the two acceptance probabilities to one and then solving for the other, \citet{barker1965monte} defined the acceptance probabilities symmetrically, sharing the total probability of acceptance between the two directions:
\[
\alpha_{\text{fwd}} + \alpha_{\text{rev}} = 1.
\]
Substituting the detailed balance condition \eqref{eq:alpha_relation} into this expression gives
\[
\text{R}\,\alpha_{\text{rev}} + \alpha_{\text{rev}} = 1,
\]
from which we obtain
\[
\alpha_{\text{rev}} = \frac{1}{\text{R}+1},
\qquad
\alpha_{\text{fwd}} = \frac{\text{R}}{\text{R}+1}.
\]
This symmetric solution also satisfies the detailed balance condition but generally yields lower acceptance probabilities than the Metropolis solution.

\renewcommand{\thesection}{C}
\setcounter{subsection}{0}

\section{Reversible Jump Markov Chain Monte Carlo in Detail}
\label{app:rjmcmc_details}

This appendix provides a detailed explanation of reversible jump Markov chain Monte Carlo (RJMCMC). The goal is to make explicit how transdimensional Metropolis--Hastings updates are constructed, why auxiliary variables and Jacobian corrections are required, and how these ingredients ensure that the resulting Markov chain targets the correct posterior distribution.

The derivations in this appendix build directly on the general theory of detailed balance and Metropolis--Hastings acceptance probabilities developed in Appendices~\ref{App::detailed_balance} and~\ref{App::metropolis_hastings}. In particular, we rely on the fact that detailed balance is guaranteed once the forward and reverse proposal densities are correctly specified on a common space and the acceptance probability is chosen according to the Metropolis rule.

\subsection{Why Standard Metropolis--Hastings Fails}

In Bayesian variable selection, the posterior distribution is defined over both model indicators $\indicators$ and model parameters $\predictors$. Different configurations of $\indicators$ correspond to parameter vectors of different dimension.

Suppose the current state is $x = (\predictors, \indicators)$, where $\predictors \in \mathbb{R}^k$, and we propose a move to $x^\star = (\predictors^\star, \indicators^\star)$, where $\predictors^\star \in \mathbb{R}^{k+1}$, corresponding to adding a predictor. Assume that the proposal is specified by a density $q(x^\star \mid x)$ on $\mathbb{R}^{k+1}$, and that the reverse move is specified by a density $q(x \mid x^\star)$ on $\mathbb{R}^k$.

For the Markov chain to have invariant distribution $p(\predictors, \indicators \mid \bm{y}, \bm{X})$, the detailed balance condition (Appendix~\ref{App::detailed_balance}) requires the pointwise equality
\begin{equation}
p(x)\, q(x^\star \mid x)\, \alpha(x \to x^\star)
=
p(x^\star)\, q(x \mid x^\star)\, \alpha(x^\star \to x).
\label{eq:DB_fail}
\end{equation}

However, Eq.~\eqref{eq:DB_fail} is not well defined. The forward proposal density $q(x^\star \mid x)$ is a density defined on $\mathbb{R}^{k+1}$, whereas the reverse proposal density $q(x \mid x^\star)$ is a density defined on $\mathbb{R}^k$.  These densities are defined on different spaces and therefore cannot be equated pointwise. As a result, the detailed balance equation cannot even be formulated, let alone solved. This is the fundamental obstruction that prevents the direct application of standard Metropolis--Hastings updates to transdimensional problems.

\subsection{Dimension Matching via Auxiliary Variables}

The central idea of RJMCMC is to construct between-model proposals in an augmented parameter space of fixed dimension. Suppose the current state of the chain is $(\predictors, \indicators)$ and we propose a new state $(\predictors^\star, \indicators^\star)$ with a different number of parameters.

To relate these states, RJMCMC introduces auxiliary variables $u$ and $u^\star$ and defines a one-to-one transformation
\[
(\predictors^\star, u^\star) = g(\predictors, u),
\]
such that
\[
\dim(\predictors^\star, u^\star) = \dim(\predictors, u).
\]
This condition is known as \emph{dimension matching}. It ensures that both the forward and reverse moves can be expressed in spaces of equal dimension, allowing the Metropolis--Hastings ratio to be constructed in the usual way.

The auxiliary variables act as placeholders for parameters that are added or removed when moving between models and facilitate the construction of reversible proposals.

Suppose the current state of the chain is $x = (\predictors, \indicators)$, and consider a between-model proposal constructed as follows. Auxiliary variables are introduced to define a one-to-one transformation between the current parameterization and that of the proposed model. Depending on the direction of the move, these auxiliary variables are either generated from a proposal distribution or obtained deterministically from the current parameters. The proposed state is then obtained through a deterministic, one-to-one transformation
\[
(\predictors^\star, u^\star) = g(\predictors, u),
\]
which defines the parameters $\predictors^\star$ of the proposed model.

\subsection{Forward and Reverse Densities on the Augmented Space}

\paragraph{Forward density expressed in the proposed coordinates.}

The between-model proposal is defined by a one-to-one transformation
\[
(\predictors^\star, u^\star) = g(\predictors, u),
\]
where $(\predictors, u)$ denotes the current parameters and auxiliary variables, and $(\predictors^\star, u^\star)$ denotes the proposed parameters and auxiliary variables. Under the forward move, the joint density
\[
p(\predictors,\indicators \mid \bm{y},\bm{X})\, q(u \mid \predictors)
\]
is defined with respect to the coordinates $(\predictors, u)$.

Because the forward density is expressed with respect to the coordinates $(\predictors,u)$, while the reverse term is evaluated with respect to $(\predictors^\star,u^\star)$, the forward density must be re-expressed in the proposed
coordinates. The proposal is defined by a one-to-one transformation $(\predictors^\star,u^\star)=g(\predictors,u)$, with inverse $(\predictors,u)=g^{-1}(\predictors^\star,u^\star)$; we write $\predictors(\predictors^\star,u^\star)$ and $u(\predictors^\star,u^\star)$ for the coordinate functions of $g^{-1}$. Expressing the forward density in these coordinates introduces a Jacobian determinant, yielding
\[
p(\predictors(\predictors^\star,u^\star),\indicators \mid \bm{y},\bm{X})\,
q(u(\predictors^\star,u^\star) \mid \predictors(\predictors^\star,u^\star))
\left|
\frac{\partial(\predictors,u)}{\partial(\predictors^\star,u^\star)}
\right|.
\]
When auxiliary variables are generated stochastically under the forward move, the proposal density is $q(u \mid \predictors)$; otherwise this density is equal to one. In all cases, the Jacobian determinant accounts for the re-expression of the forward density in the proposed coordinates.

\paragraph{Reverse density induced by the inverse transformation.}

Under the present construction, the reverse state $(\predictors,\indicators,u)$ is recovered deterministically from $(\predictors^\star,\indicators^\star,u^\star)$ via the inverse transformation
\[
(\predictors,u) = g^{-1}(\predictors^\star,u^\star).
\]
Because the reverse density is already defined with respect to the proposed coordinates
$(\predictors^\star,u^\star)$, no change of variables is required. Accordingly, the reverse density takes the form
\[
p(\predictors^\star,\indicators^\star \mid \bm{y},\bm{X})\,
q(u^\star \mid \predictors^\star).
\]
When the reverse move involves stochastic generation of auxiliary variables, the proposal density is $q(u^\star \mid \predictors^\star)$; otherwise it is equal to one.

\subsection{Metropolis--Hastings Acceptance Ratio}

The Metropolis-Hastings ratio combines the posterior density ratio, the auxiliary-variable proposal densities, and the Jacobian determinant induced by the transformation on the augmented space.
\[
\frac{
p(\predictors^\star,\indicators^\star \mid \bm{y}, \bm{X})\,
q(u^\star \mid \predictors^\star, \indicators^\star)
}{
p(\predictors,\indicators \mid \bm{y}, \bm{X})\,
q(u \mid \predictors, \indicators)
}
\left|
\frac{\partial(\predictors^\star, u^\star)}
{\partial(\predictors, u)}
\right|.
\]
Applying the Metropolis acceptance rule to this ratio yields the RJMCMC acceptance probability.

\renewcommand{\thesection}{D}
\setcounter{subsection}{0}
\section{Adaptive Metropolis-Hastings via Robbins-Monro Updates}
\label{App::adaptive_metropolis}

The efficiency of a random walk Metropolis algorithm depends critically on the scale of the proposal distribution. Proposals that are too diffuse lead to low acceptance rates, whereas proposals that are too concentrated result in slow exploration of the posterior despite frequent acceptance. Optimal scaling results suggest that algorithm efficiency is maximized when the acceptance rate is close to a problem-dependent target value \citep{RobertsGelmanGilks1997, RobertsRosenthal_2001}.

Adaptive tuning of proposal variances can be viewed as a stochastic approximation problem in which the proposal scale is adjusted to target a desired acceptance rate. Robbins-Monro algorithms provide a principled framework for solving such problems when only noisy evaluations of the acceptance probability are available \citep{RobbinsMonro_1951}.

Let $\tau_i^{(t)}$ denote the proposal standard deviation for coefficient $\predictor_i$ at iteration $t$ of the warmup phase. Proposals are generated according to
\[
\predictor_i^\ast \sim \mathcal{N}\left(\predictor_i, \tau_i^{(t)\,2}\right).
\]
After evaluating the Metropolis-Hastings acceptance step, the proposal standard deviation is updated on the logarithmic scale according to
\begin{equation}
\label{eq:rm_update}
\log\bigl(\tau_i^{(t+1)}\bigr) =
\log\bigl(\tau_i^{(t)}\bigr) + \frac{1}{(t+1)^\phi} \left( \mathbf{1}\{\text{accept}\} - \alpha_{\mathrm{target}} \right),
\end{equation}
where $\mathbf{1}\{\text{accept}\}$ equals one if the proposal is accepted and zero otherwise, and $\alpha_{\mathrm{target}}$ denotes the target acceptance probability.

This update is a Robbins-Monro iteration that seeks the root of the equation
\[
\mathbb{E}[\mathbf{1}\{\text{accept}\}] = \alpha_{\mathrm{target}},
\]
which adjusts the proposal variance so that the long-run acceptance rate matches the desired target. Updating $\tau_i$ on the logarithmic scale ensures positivity of the proposal variance and yields multiplicative rather than additive adjustments.

The step-size sequence $(t+1)^{-\phi}$ satisfies the standard Robbins-Monro conditions provided that $\phi \in (1/2,1]$, ensuring diminishing adaptation and stability of the updates \citep{RobbinsMonro_1951}. We use $\phi = 0.75$, which provides a balance between rapid initial adaptation and stable convergence of the proposal variances.

In the add–delete scheme for variable selection under MoMS, a single component of $\predictors$ is added or removed at each step, leading to one-dimensional random-walk Metropolis updates.  Following theoretical and empirical results for componentwise random-walk Metropolis algorithms, we set the target acceptance rate to $\alpha_{\mathrm{target}} = 0.44$ for these univariate updates \citep{RobertsRosenthal_2001}.

\subsection{Implementation for the linear regression model}

Adaptation is restricted to the warmup phase of the Markov chain, after which the proposal variances are fixed to preserve the Markov property and ensure correct stationary behavior of the sampler. During warmup, we assume the full model $\mathcal{M}_{\ast}$ and use adaptive Metropolis updates to learn the proposal variances $\tau_i$ for all coefficients. During the sampling phase, the proposal variances are fixed, and between-model moves use random-walk Metropolis updates, while within-model updates of both $\predictors$ and $\sigma^2$ are performed using Gibbs sampling.

%\renewcommand{\thesection}{E}
%\setcounter{subsection}{0}
%\section{Advanced Between-Model Moves}\label{App:AdvancedMoves}

\renewcommand{\thesection}{E}
\setcounter{subsection}{0}\section{Effective Sample Size for Inclusion Indicators}
\label{App::ess}

To compare the efficiency of MoMS and RJMCMC, a natural metric is the effective sample size. However, common implementations assume that the sampled parameters are continuous rather than binary. In variable selection problems, inference about inclusion is based on a discrete indicator sequence, and slow switching between models can induce strong autocorrelation. \citet{HeckEtAl_2019} propose quantifying this dependence by modeling the sequence of visited models as a discrete Markov chain. We instead consider the marginal inclusion indicator for a single effect and model it as a two-state Markov chain. From this representation we derive the effective sample size (ESS) and expressions for the Monte Carlo standard error (MCSE).

Let $\{\indicator_t\}$ be a binary MCMC chain where $\indicator_t \in \{0, 1\}$ indicates whether a parameter is included (1) or excluded (0) at iteration $t$. We model this sequence as a homogeneous first-order Markov chain, meaning that 
\[
    P(\indicator_{t+1} \mid \indicator_t, \indicator_{t-1}, \ldots) = P(\indicator_{t+1} \mid \indicator_t)
\]
and that the transition probabilities do not depend on $t$ \citep{Tierney_1994_MarkovChains}.

The transition matrix is 
\[
\mathbf{P} =
\begin{pmatrix}
1 - a & a \\
b & 1 - b
\end{pmatrix}
\]
where $a = P(\indicator_{t+1} = 1 \mid \indicator_t = 0)$ is the probability of switching from $0$ to $1$, and $b = P(\indicator_{t+1} = 0 \mid \indicator_t = 1)$ is the probability of switching from $1$ to $0$.

The stationary distribution $\boldsymbol{\pi} = (\pi_0, \pi_1)$ satisfies $\boldsymbol{\pi} \mathbf{P} = \boldsymbol{\pi}$ and $\pi_0 + \pi_1 = 1$. Solving the balance equation $\pi_0 a = \pi_1 b$ yields
\begin{align*}
\pi_0 &= \frac{b}{a + b},\\ 
\pi_1 &= \frac{a}{a + b}.
\end{align*}

To compute the lag-$k$ autocorrelation, note that the eigenvalues of $\mathbf{P}$ are $1$ and $1 - a - b$ \citep[see Section 5.5,][]{Privault2013}. The lag-$k$ autocorrelation is therefore
\[
\rho(k) = (1 - a - b)^k.
\]

The integrated autocorrelation time is 
\[
\tau_{\mathrm{int}}=1 + 2\sum_{k=1}^\infty \rho(k).
\]
Because this is a geometric series,
\[
\tau_{\mathrm{int}}=\frac{2 - (a + b)}{a +b}.
\]
The effective sample size for a chain of length $T$ is therefore
\[
    \mathrm{ESS} = \frac{T}{\tau_{\mathrm{int}}} = T\frac{a +b}{2 - (a + b)}.
\]

In practice, $a$ and $b$ are unknown and must be estimated from the observed transition counts. Let $n_{ij}$ denote the number of transitions from state $i$ to state $j$, for $i, j \in \{0, 1\}$. 
Then
\begin{align*}
\hat{a} &= \frac{n_{01}}{n_{00} + n_{01}}, \\
\hat{b} &= \frac{n_{10}}{n_{10} + n_{11}}.
\end{align*}
 
The Monte Carlo standard error (MCSE) for the estimated inclusion probability $\hat{p}$ follows directly:
\[
 \mathrm{MCSE}(\hat{p}) = \sqrt{\frac{\mathrm{Var}(\indicator_t)}{\mathrm{ESS}}} = \sqrt{\frac{\hat{p} (1 - \hat{p})}{\mathrm{ESS}}}.
\]

%\renewcommand{\thesection}{F}
%\setcounter{subsection}{0}\section{Empirical Example}

%\begin{table}[!ht]
%\centering
%\caption{The 20 DRM wordlist used in \citep{mak2023registered}.}
%\label{tb:lureTable}
%\include{tables/lureTable.tex}
% \tablenote{}
%\end{table}

\renewcommand{\thesection}{G}
\setcounter{subsection}{0}\section{Empirical Example}

% \begin{figure}[!ht]
%     \centering
%     \caption{Correlations among Factor Loadings in the Multidimensional Generalized Partial Credit Model}
%     \includegraphics[width=\linewidth]{Figures/factor_loadings_posterior_correlation.pdf}
%     \label{fig:gpcm_correlation}
%     \figurenote{The left panel shows the posterior means of the first two dimensions for each items. Point color indicates the posterior correlation between the corresponding loadings. The magnitude of the correlation increases with the magnitude of the posterior means; correlations are positive when the loadings share the same sign and negative when their signs differ. Item 20 is highlighted with a square; the right panel displays the corresponding bivariate posterior.}
% \end{figure}

\begin{figure}[!ht]
    \centering
    \caption{Shrinkage due to Variable Selection on the Norm of $\bm{\alpha}$}
    \includegraphics[width=\linewidth]{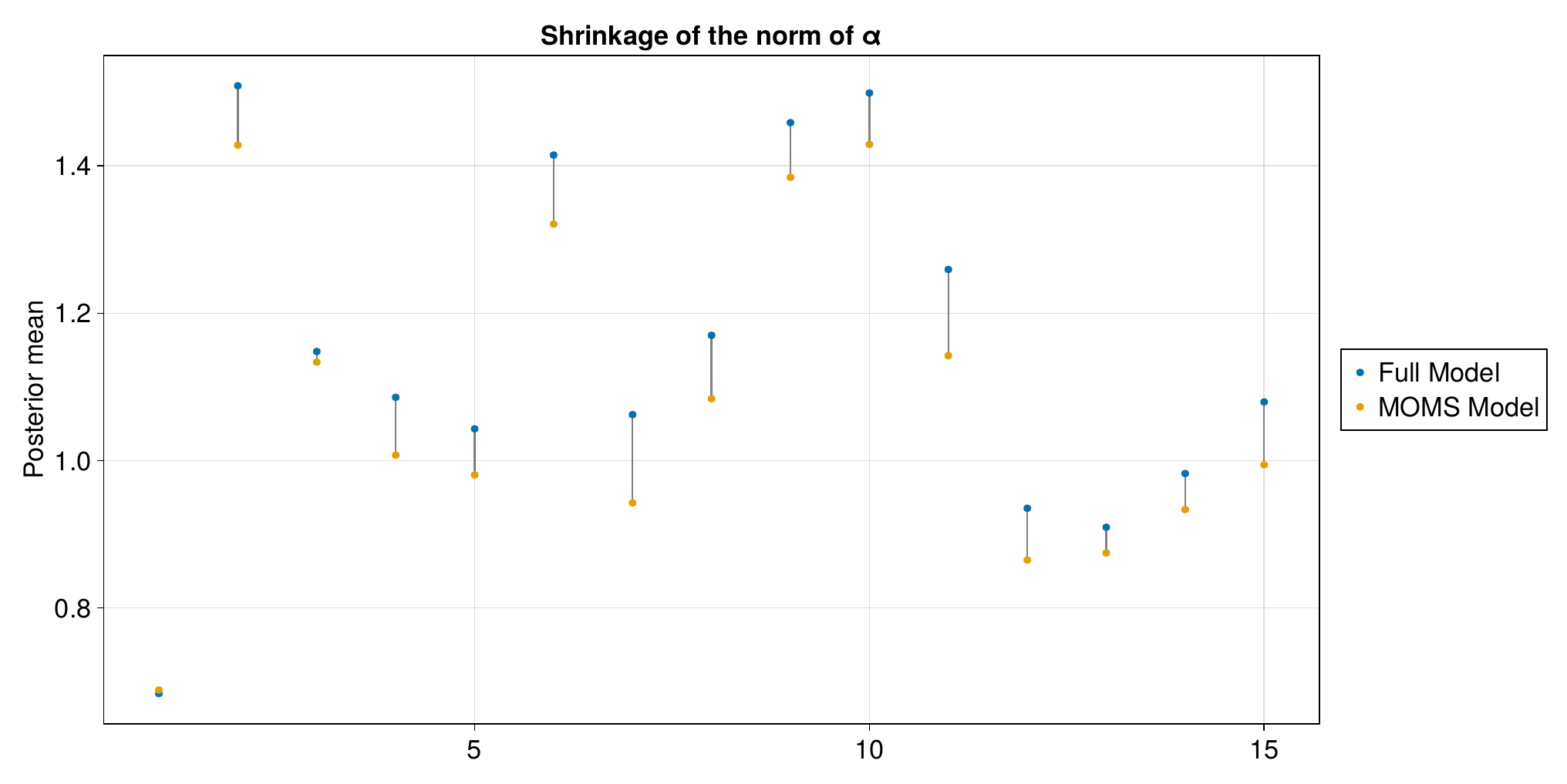}
    \label{fig:gpcm_norm_shrinkage}
\end{figure}
\end{document}